\documentclass[manuscript]{aa}

\usepackage{graphicx}
\usepackage[varg]{txfonts}
\usepackage{hyperref}
\usepackage{xcolor}
\usepackage{soul}

\usepackage{multirow}
\usepackage{amsmath}
\usepackage{hyperref}
\usepackage{booktabs}
\usepackage{lipsum}
\usepackage{ulem}

\newcommand\mancha{\textsc{Mancha3D}}

\newcommand{\qrad}{Q$_{\rm rad}$}
\newcommand{\qloss}{Q$_{\rm loss}$}
\newcommand{\qtc}{{\bf q}} 

\newcommand{\qradEQ}{\text{Q}_\text{rad}}
\newcommand{\qlossEQ}{\text{Q}_\text{loss}}

\hypersetup{
    colorlinks=true,       
    linkcolor=blue,        
    citecolor=blue,       
    urlcolor=blue           
}

\begin{document}

\title{Modeling Solar Atmosphere Dynamics with MAGEC{} }

\authorrunning{A. Navarro et al.}

\author{Anamar\'ia Navarro\and 
E. Khomenko\and 
N. Vitas \and 
T. Felipe
}

\institute{Instituto de Astrof\'isica de Canarias, 38205 La Laguna, Tenerife, Spain 
\and
Dpto. de Astrof\'isica, Universidad de La Laguna, 38205 La Laguna, Tenerife, Spain \email{anavarro@iac.es} 
}

\date{\today}



\abstract{
Modeling the solar atmosphere is challenging due to its layered structure and dynamic multi-scale processes.
}
{
We aim to validate the new radiative MHD code MAGEC—built by integrating the \mancha{} and MAGNUS codes into a finite-volume, shock-capturing framework—and to explore its capabilities through 2D simulations of magneto-convection in the solar atmosphere.
}
{
The MAGEC code is parallelized with MPI, enabling efficient scalability for large-scale simulations. We have enhanced it with advanced numerical techniques to address the specific complexities of the solar corona, including a module for LTE radiative coronal losses. To address the small time steps due to large heat flux values, we adopted the hyperbolic treatment for the thermal conduction of \mancha{}, which significantly improves the computational times.  In addition, we estimated the effective numerical resistivity and viscosity through a dedicated set of experiments. To evaluate the robustness and accuracy of MAGEC, we performed a series of 2D simulations covering a domain extending from 2 Mm below the solar surface to 18.16 Mm into the corona. Simulations were conducted with both open and closed magnetic field configurations.   For each case, we analyzed the resulting steady-state temperature profiles and examined the energy contributions at different heights. In addition, we investigated the influence of the perpendicular component of thermal conduction in a dedicated simulation.

}
{
The MAGEC code effectively reproduced expected temperature profiles based on the boundary conditions applied and the imposed magnetic field configuration. All simulations reached a thermally stable state. When using an open vertical magnetic field, the temperature in the middle corona was higher than in the case with a closed, arcade-like magnetic field structure. We quantified the contributions to the internal energy from all explicit and implicit terms in the steady state, both in terms of temporal averages and as functions of height, as well as their relative contributions to total heating and cooling. In a second phase of the study, we investigated the role of the perpendicular component of thermal conduction, often neglected in coronal models, and found that it can influence plasma dynamics around reconnection events. Although local effects are modest, their cumulative impact can lead to measurable changes in the average temperature profile.

}
{
Through detailed validation, MAGEC is a reliable and efficient code for radiative MHD simulations of the solar atmosphere. The integration of shock-capturing methods is particularly well-suited to modeling the plasma environment, effectively handling the shocks and discontinuities characteristic of the solar atmosphere. MAGEC is a robust tool for high-fidelity magneto-convection simulations of the solar atmospheric dynamics.
}

\maketitle

\section{Introduction}\label{sec:intro}

Modeling the solar atmosphere presents a significant computational challenge because of the complexity and diversity of its structure. The solar atmosphere comprises multiple layers—each with distinct dynamics, physical processes, and scales—that interact intricately with each other. From the convective motions in the photosphere to the highly structured magnetic fields in the corona, each layer requires specific modeling techniques to accurately capture its unique characteristics. Developing a code that can address these diverse and dynamic processes is a complex task. As our understanding of solar physics deepens and computational capabilities advance, it becomes crucial to continuously improve and adapt numerical codes to better capture the complexities inherent in the solar atmosphere. 

The simulation of the solar atmosphere has evolved considerably over the past few decades, with successive studies incorporating increasingly sophisticated physics and numerical methods. Several representative works have focused on the layers up to the photosphere. For example, \citet{1998ApJ...499..914S} investigated solar granulation using the STAGGER code, emphasizing convective motions and energy transport in the absence of magnetic fields. \citet{1998ApJ...495..468S} explored the interaction between magnetic flux sheets and granular convection, highlighting dynamic phenomena such as shock formation, using an early version of the CO5BOLD code. \citet{2005A&A...429..335V} developed the MURAM code to simulate solar magneto-convection, analyzing the concentration and amplification of magnetic flux. \citet{2006ApJ...642.1246S} used STAGGER to study small-scale magneto-convection and the evolution of magnetic flux. Furthermore, \citet{2008ApJ...684L..51J}, using the SOLARBOX code, examined the influence of magnetic fields on convection and acoustic oscillations.

Further studies have extended into the chromosphere, where additional complexities arise from this layer's partial ionization and strong interactions with radiation. \citet{2005ESASP.596E..65S} introduced the CO5BOLD code to model the solar atmosphere from the convection zone to the chromosphere, focusing on dynamic magnetic flux reshuffling, shock propagation, and flux tube formation. Using an earlier version of the BIFROST code \citep{bifrost_code}, \citet{2010MmSAI..81..582C} performed 3D radiative MHD simulations to investigate chromospheric heating, emphasizing mechanisms such as Ohmic dissipation, magnetic field interactions, and swirling motions as channels for energy transport. More recently, \citet{2018A&A...618A..87K} employed the \mancha{} code \citep{2024SoPh..299...23M} to study the role of ambipolar diffusion in wave propagation, Poynting flux absorption, and chromospheric heating, highlighting the importance of ion-neutral interactions in the partially ionized chromospheric plasma.

Increasing the complexity of solar models, several studies have succeeded in including the solar corona within the same computational domain as the convection zone and chromosphere. Among the first to achieve this were \citet{2007ApJ...665.1469A}, using the RADMHD code, and \citet{2007ASPC..368..107H}, who employed 3D MHD simulations with the STAGGER code to model the dynamic coupling between these layers. These pioneering works highlighted the role of magneto-convection in driving energy and magnetic flux transport and emphasized the importance of magnetic connectivity in shaping coronal structures and enabling energy dissipation for heating.

Later, \cite{2017ApJ...834...10R} extended the MURaM code to simulate the corona, incorporating key physical processes such as optically thin radiative losses, field-aligned thermal conduction, and the Boris correction to manage high Alfvén speeds. Similarly, the BIFROST code \citep{bifrost_code} - which includes non-gray, non-LTE (non-local thermodynamic equilibrium) radiative transfer, field-aligned thermal conduction, ambipolar diffusion, and the Hall effect - was employed by \citet{2017ApJ...847...36M, 2020ApJ...889...95M} to investigate the influence of ion-neutral interactions and non-equilibrium ionization on chromospheric heating and spicule dynamics. 

The two-fluid JOANA code \citep{2018MNRAS.481..262W} has been widely used for many solar applications; for example, \citet{2022Ap&SS.367..111M} demonstrated how granulation-generated waves and plasma viscosity can balance radiative losses in the chromosphere. The MAGNUS code, derived from earlier work by \citet{magnus} and built on the CAFE \citep{CafeRelativista} and Newtonian-CAFE \citep{CafeNewtoniano} codes, has been applied to model chromospheric jets \citep{2021MNRAS.500.3329N} and flux emergence \citep{2019MNRAS.489.2769N}.

The RAMENS code was used by \citet{2023ApJ...951L..47I} to simulate the formation of the solar wind, highlighting the roles of convection-driven wave excitation and magnetic reconnection. The R2D2 code was used by \citet{2024ApJ...975..209T} to study the impact of magnetic flux tube twist on flux emergence, revealing key processes underlying active region formation. A wide range of applications in solar and astrophysical plasma modeling have also been carried out using codes such as AMRVAC \citep{2023A&A...673A..66K}, LARE3D \citep{2001JCoPh.171..151A}, PENCIL \citep{2021JOSS....6.2807P}, PLUTO \citep{pluto_code}, and GOEMHD3 \citep{2015A&A...580A..48S}.

Expanding into stellar contexts, \citet{2024LRCA...10....2C} reviewed simulations performed with CO5BOLD \citep{2013MSAIS..24...26F, 2017MmSAI..88...12F} and Athena++ \citep{athena_code}, uncovering the global-scale nature of convection in cool, evolved stars and its significant role in driving stellar winds and mass loss. Similarly, \citet{2022A&A...664A..24C} used the RAMSES code to investigate stellar magneto-convection. Incorporating adaptive mesh refinement techniques for large-scale solar simulations, the codes DISPATCH \citep{2025A&A...698A..69P} and DYABLO \citep{2025JPhCS2997a2014D} are being developed.

These codes adopt a variety of numerical methods. Some use finite-volume methods with shock-capturing schemes such as CO5BOLD, RAMSES, MAGNUS, JOANA, DISPATCH, and DYABLO. The rest employ finite-difference methods, and some have modular frameworks such as AMRVAC and PLUTO that support both schemes. Among the codes that implement shock-capturing schemes, only CO5BOLD, JOANA, DISPATCH, and DYABLO have been used to simulate magneto-convection. Of these, only JOANA has employed a simulation domain that extends into the corona and includes key coronal physics, such as radiative losses and thermal conduction.

Finite-volume methods with high-resolution shock-capturing techniques are particularly well suited for simulations of the solar atmosphere, as they are designed to handle complex fluid dynamics. Unlike traditional numerical schemes, advanced approaches such as Godunov-type methods use approximate Riemann solvers to accurately capture shocks and discontinuities \citep{Toro}. These techniques allow the resolution of sharp gradients and abrupt changes in the plasma state without relying on artificial diffusivity, which can obscure the underlying physics. However, their implementation is generally more complex than finite-difference schemes, and they may require more computational resources, especially in multi-dimensional setups.

In this paper, we present the new numerical code MAGEC, which takes advantage of shock-resolving finite volume techniques. The paper is organized as follows: Section \ref{sec:MAGEC} provides a detailed introduction to the newly developed MAGEC code. The fundamental equations governing the simulations are outlined in Section \ref{subsec:equations}, followed by a description of the numerical schemes implemented in Section \ref{subsec:num_methods}. The thermal conduction model and its implementation are discussed in Section \ref{subsec:tc}, while the equation of state is presented in Section \ref{subsec:eos}. Section \ref{subsec:rad} gives an overview of the radiative modules, and Section \ref{subsec:numerical_resistivity_viscosity} describes the calculation of the numerical resistivity and viscosity. The simulation setup and results are presented in Section \ref{sec:Simulations}. Finally, Section \ref{sec:conclusions} summarizes the main findings and a discussion for potential future research.


\section{Description of the MAGEC code} \label{sec:MAGEC}

The MAGEC code is a finite volume-based tool developed to model magnetohydrodynamic (MHD) phenomena in the solar atmosphere and other astrophysical environments. It employs shock-capturing methods for high-precision simulations. MAGEC integrates the \mancha{} code \citep{2024SoPh..299...23M} with the MAGNUS code \citep{magnus}, merging the advanced numerical schemes of  MAGNUS with the efficient, MPI-parallelized framework of  \mancha{} and its physical modules. 

Beyond incorporating \mbox{MAGNUS's} core schemes, MAGEC includes critical modules from \mancha{} for optically thick LTE radiative transfer, thermal conduction, and non-ideal physics derived from plasma partial ionization in the single-fluid approximation (ambipolar diffusion and Hall effects), though the latter features are not applied in the present study. The code also integrates a newly developed module for coronal thin radiative losses. This section offers a comprehensive overview of each component within the code.

\subsection{Equations} \label{subsec:equations}

A gravitationally stratified plasma under the influence of thermal conduction and radiative terms can be represented by the following set of equations: 
\begin{align}
& \frac{\partial \rho}{\partial t} + \nabla \cdot \left( \rho {\bf v} \right) = 0 \, , \label{eq:rho} \\
 & \frac{\partial{\bf B}}{\partial t} + \nabla \cdot \left( {\bf v} {\bf B} - {\bf B} {\bf v} \right) = 0 \, , \label{eq:B}  \\
 & \frac{\partial (\rho {\bf v})}{\partial t} + \nabla \cdot \left[   \rho {\bf v} {\bf v} + \left( p + \frac{{\bf B}^2}{2\mu_0} \right) {\bf I} - \frac{{\bf B B} }{\mu_0}  \right] = \rho {\bf g} \, ,  \label{eq:rhoV}  \\
 & \frac{\partial e}{\partial t} + \nabla \cdot \left[   \left(  e + p + \frac{{\bf B}^2}{2 \mu_0} \right){\bf v}   - \frac{\bf B}{\mu_0} \left(  {\bf B \cdot v} \right)   \right]  = \label{eq:energy}  \\
 & \hspace{3.5 cm} \rho {\bf v} \cdot {\bf g} - \nabla \cdot {\bf q} +  \qradEQ + \qlossEQ   \, , \nonumber
\end{align}
where $\rho$ is the density, ${\bf v}$ is the velocity, $p$ is the gas pressure, ${\bf B}$ is the magnetic field, ${\bf g}$ is the gravitational acceleration, $\bf{q}$ is the heat flux vector, \qrad/\qloss{} are the radiative terms in optically thick/thin plasma, and $\mu_0$ is the magnetic permeability of free space. The dot ``$\cdot$'' represents the scalar product of vectors, while the notation ``{\bf B}{\bf B}'' stands for the tensor product and ${\bf I}$ is the identity tensor. In equation \eqref{eq:energy} $e$ represents the total energy per unit volume given by 
\begin{equation}
e = e_{\text{int}} + \frac{1}{2} \rho v^2 + \frac{\mathbf{B}^2}{2\mu_0}, 
\end{equation}
where $e_{\text{int}}$ is the internal energy per unit volume given by the equation of state, which will be discussed in subsection \ref{subsec:eos}.

\subsection{Numerical methods} \label{subsec:num_methods}

The same system of equations \eqref{eq:rho}-\eqref{eq:energy} can be written in the conservative form
\begin{eqnarray}
\frac{\partial \mathbf{U}}{\partial t} + \nabla \cdot     \mathbb{F} = \mathbf{S} \, , \label{eq:conservative_form}
\end{eqnarray}
where $\mathbf{U}$ is the vector of conservative variables, $\mathbb{F}$ is the flux dyad and $\vec{S}$ is the vector of source/sink terms given by $\mathbf{U} = \left[\rho, \rho \mathbf{v},e, \mathbf{B} \right]^T$,
\begin{eqnarray}
\mathbb{F} = 
\begin{bmatrix}
\rho {\bf v} \\
\rho {\bf v} {\bf v} + \left( p + \dfrac{\mathbf{B}^2}{2\mu_0} \right) \mathbf{I} - \dfrac{\mathbf{B} \mathbf{B}}{\mu_0} \\
\left(e + p\right) \mathbf{v} - \dfrac{\mathbf{B}}{\mu_0} \left( \mathbf{B} \cdot \mathbf{v} \right) \\
\mathbf{v} \mathbf{B} - \mathbf{B} \mathbf{v} 
\end{bmatrix},
\end{eqnarray}
and
\begin{eqnarray}
\mathbf{S} = 
\begin{bmatrix}
0 \\
\rho \mathbf{g}  \\
-\nabla \cdot \mathbf{q} + \qradEQ + \qlossEQ  \\
0
\end{bmatrix} \, .   
\end{eqnarray}

The equations are solved using the method of lines, which transforms PDEs into ODEs via spatial discretization with a finite volume (FV) approach, followed by time integration,
\begin{eqnarray}
 \dfrac{ \mathrm{d} \mathbf{U}_{(i,j,k)}}{ \mathrm{d} t} =
 - \frac{\mathbb{F}^x_{(i+1/2,j,k)} - \mathbb{F}^x_{(i-1/2,j,k)}}{\Delta x}  - \frac{\mathbb{F}^y_{(i,j+1/2,k)} - \mathbb{F}^y_{(i,j-1/2,k)}}{\Delta y}  \\
  - \frac{\mathbb{F}^z_{(i,j,k+1/2)} - \mathbb{F}^z_{(i,j,k-1/2)}}{\Delta z}
+ S_{(i,j,k)}\, , \hspace{2cm}   \nonumber   
\end{eqnarray}
where $\mathbb{F}^x_{(i \pm 1/2,j,k)}$, $ \mathbb{F}^y_{(i,j \pm 1/2,k)} $, and $\mathbb{F}^z_{(i,j,k \pm 1/2)}$ are the numerical fluxes at the interfaces of a cell and will be calculated using High-Resolution Shock Capturing methods (HRSC), which couple slope limiters and Riemann solvers.

The modular structure of the code allows for a flexible combination of numerical methods. For time integration, the code includes third-order Strong-Stability preserving Runge-Kutta time-steppers (SSP-RK3) \citep{gottlieb2001ssp} as well as fourth- and sixth-order standard Runge–Kutta schemes. The available HRSC methods include the Riemann solvers HLL \citep{Harten_etal_1983}, HLLC \citep{LI2005344}, and HLLD \citep{MIYOSHI2005315} and the slope limiters: MINMOD \citep{minmod}, MC \citep{VANLEER1977263}, Van Leer \citep{VANLEER1974361}, Ospre \citep{ospre}, and WENO5 \citep{TitarevToro2004}. The code uses the Flux Constrained Transport method \citep{CT_evans} to prevent the divergence of magnetic field from growing in time due to numerical errors. The general expressions of the methods are provided in the appendix \ref{appendix}.

The time step due to advection is 
\begin{equation}
\Delta t = C_{\text{cfl}} \times \mathrm{min}_{ijk} \left( \frac{\Delta x}{|\lambda^{x}_{ijk}|}, \frac{\Delta y}{|\lambda^{y}_{ijk}|}, \frac{\Delta z}{|\lambda^{z}_{ijk}|} \right) \, ,
\end{equation}
where $C_\text{cfl}$ stands for the Courant number, and $\lambda^{d}_{ijk}$ is the speed of the fastest wave present traveling in direction $d=x,y,z$ at the grid cell indexed by $(i,j,k)$. The minimum is taken over all grid cells of the domain. The values of $\lambda^{d}_{ijk}$ are obtained from the eigenvalues of the Jacobian matrix associated with the system of equations \eqref{eq:conservative_form}.  For instance, in the $x-$direction, they are given by 
\begin{eqnarray}
\lambda_1 = v_x             \, , \ \ 
\lambda_{2,3} = v_x \pm v_a \, , \ \
\lambda_{4,5} = v_x \pm c_s \, , \ \
\lambda_{6,7} = v_x \pm c_f \, , \ \ \label{eq:eigenvalues}
\end{eqnarray}
where $v_a$ is the Alfvén speed and $c_f$ and $c_s$ are the fast and slow magnetosonic speeds given by
\begin{eqnarray}
v_a =  \dfrac{B_x}{\sqrt{\mu_0 \rho}} \, , \ \ 
c_{f,s} =  \dfrac{1}{\sqrt{2}} \sqrt{ a^2 + v_a^2  \pm \sqrt{\left(a^2 + v_a^2\right)^2 - 4 a^2 v_a^2} } \, , \label{eq:va_and_cfs}
\end{eqnarray}
with $a$ the speed of sound.

\subsection{Thermal conductivity} \label{subsec:tc}

The effects of thermal conduction (TC) are represented by the term $\nabla \cdot \mathbf{q}$ in the energy equation (\ref{eq:energy}). In the strongly magnetized regime, the heat flux is aligned with the magnetic field,
\begin{eqnarray}
{\bf q}= - \kappa_{\|} \nabla_{\|}T \, ,
\end{eqnarray}
where $\nabla_{\|}$ gives the projection in the parallel direction to the magnetic field and $\kappa_{\|}$ is the thermal conductivity. This approximation is valid for most astrophysical applications and is commonly referred to as Spitzer’s \citep{Spitzer1956} thermal conduction. 

If the magnetization is not strong, the propagation of heat along the other directions can be expressed in the following form
\begin{eqnarray}
{\bf q}= - \kappa_{\|} \nabla_{\|}T - \kappa_{\perp}\nabla_{\perp}T +\kappa_{\times} {\bf \hat{b}} \times \nabla_{\perp} T \, ,
\end{eqnarray}
where $\nabla_{\perp} = \nabla - \nabla_{\|}$ is the projection operator in the perpendicular direction to the magnetic field, and the last term is the projection in the transverse direction (second perpendicular direction to the magnetic field). \cite{1965RvPP....1..205B} deduced the general expressions for the conductivity coefficients ($\kappa_{\|}$, $\kappa_{\perp}$ and $\kappa_\times$) for electrons and ions in terms of the plasma properties like density, temperature, pressure and magnetic field.
In the case of electrons, they are given by
\begin{align}
\kappa_\parallel^e &= \frac{3.1616 \, k_B p_e}{\nu_{ei} m_e},   \label{eq:kappa_par_e} \\ 
\kappa_\perp^e &= \frac{k_B p_e}{\nu_{ei} m_e} \frac{4.664 x_e^2 + 11.92}{x_e^4 + 14.79 x_e^2 + 3.77},  \label{eq:kappa_perp_e} \\
\kappa_\times^e &= \frac{k_B p_e}{\nu_{ei} m_e} \frac{\frac{5}{2} x_e^2 + 21.67}{x_e^4 + 14.79 x_e^2 + 3.77},  \label{eq:kappa_times_e}
\end{align} 
and for ions,
\begin{align}
\kappa_\parallel^i &= \frac{3.906 \, k_B p_i}{\nu_{ii} m_i},    \label{eq:kappa_par_i} \\
\kappa_\perp^i &= \frac{k_B p_i}{\nu_{ii} m_i} \frac{2 x_i^2 + 2.645}{x_i^4 + 2.70 x_i^2 + 0.677},  \label{eq:kappa_perp_i} \\
\kappa_\times^i &= \frac{k_B p_i}{\nu_{ii} m_i} \frac{\frac{5}{2} x_i^2 + 4.65}{x_i^4 + 2.70 x_i^2 + 0.677}.  \label{eq:kappa_times_i} 
\end{align}
where the subscripts and superscripts  $e$ and $i$ denote the electrons and ions, respectively. The pressures and masses of the two species are represented by $p_e$, $p_i$, $m_e$, and $m_i$. The collision frequencies are given as $\nu_{ii}$ for ion-ion collisions and $\nu_{ei}$ for electron-ion collisions. The variables $x_i$ and $x_e$  represent the ratio of the cyclotron frequency ($\Omega$) to the collision frequency ($\nu$) for each specie, respectively
\begin{align}
x_e &= \frac{\Omega_e}{\nu_{ei}}, \quad x_i = \frac{\Omega_i}{\nu_{ii}}, 
\end{align}
where
\begin{align}
\Omega_e &= \frac{|e|B}{m_e}, \quad \nu_{ei} = 3.7 \times 10^{-6} \frac{\ln(\Lambda)n_e}{T_e^{3/2}}, \quad p_e = n_e k_B T_e, \\
\Omega_i &= \frac{|e|Z_i B}{m_i}, \quad \nu_{ii} = 6 \times 10^{-8} \frac{\ln(\Lambda)n_i Z_i^4}{T_i^{3/2}}, \quad p_i = n_i k_B T_i, 
\end{align}
$e$ is the electron charge, $Z_i$ is the charge of the ion, $n_e$ and $n_i$  are the electron and ion number densities. All definitions are in SI units. The Coulomb logarithm can be approximated by:
\begin{align}
\ln(\Lambda) = 23.4 - 1.15 \log_{10}(n_e) + 3.45 \log_{10}(T_e), 
\end{align}
for $ T_e < 50$eV, with $n_e$ given in cm$^{-3}$ and $T_e$ in electronvolts. From these expressions, we can see that in the case of a null magnetic field, the heat flux becomes isotropic $\mathbf{q} = -\kappa_\parallel \nabla T $ since $x_e = 0 $ and $ \kappa_\perp^e = \kappa_\parallel^e $. 

The parallel conductivity values predicted by the two models are nearly identical, with the value in the Spitzer model being 1.013094636 times that predicted by the Braginskii model. The difference between the models lies in their treatment of conduction: the Spitzer model considers only conduction along the magnetic field, whereas the Braginskii model accounts for propagation in all directions.

The numerical treatment of thermal conduction is challenging in regions with high temperatures and strong magnetic fields. Under these conditions, the parallel component of the heat flux becomes too high, causing a reduction of the timestep associated with heat conduction. To overcome this difficulty, we use the hyperbolic heat-conduction equation, which allows for larger timesteps.

This equation modifies the classical Fourier heat flux by adding a thermal inertia term. It was independently proposed by \citet{vernotte1958paradoxes} and \citet{cattaneo1958forme} to resolve the paradox of infinite propagation speed in Fourier’s theory, leading to the concept of thermal waves. This approach has been extensively studied both theoretically and experimentally (see \citealt{ABDELHAMID1999899} and references therein). In astrophysics, it was first applied to model field-aligned diffusion of cosmic rays by \citet{2006MNRAS.373..643S} and later introduced in solar simulations by \citet{2017ApJ...834...10R} in the MURAM code. It has since been implemented in the PENCIL code \citep{2020GApFD.114..261W}, the MANCHA code \citep{2022A&A...663A..96N}, and the BIFROST code \citep{2024A&A...692A.139C}. 

This method introduces an additional hyperbolic equation to evolve the parallel component of the heat flux $q_{\|}$, the equation is given by
\begin{eqnarray}
\frac{\partial q_{\|}}{\partial t} = \frac{1}{\tau} \left( -  f_{\text{sat}} \kappa_{\|} \left( {\bf \hat{b}} \cdot \nabla \right) T - q_{\|} \right) \, .
\end{eqnarray}
where $f_{\text{sat}}$ is the saturation factor used for stability   
\begin{align}
f_{\text{sat}} = \left( 1 + \frac{|\kappa_{\|} (\mathbf{\hat{b}} \cdot \nabla) T|}{1.5 \rho c_s^3} \right)^{-1},
\end{align}
and $\tau$ is the relaxation time, which should be calibrated depending on the problem. For the simulations used in this paper we used a value of $\tau = 10 \Delta t$.  

In the code, the user can select the thermal conduction model appropriate for their simulation: either the Braginskii or Spitzer formulation for solar plasma conditions, or, for more general MHD applications, specify constant conductivity values manually. Additionally, the user can choose between a hyperbolic or parabolic treatment.

\subsection{Equation of state} \label{subsec:eos}

The system of equations is closed using an equation of state (EOS) for the solar chemical composition, as provided by \cite{1989GeCoA..53..197A}. The internal energy per unit volume, density, and electron pressure are precomputed over a temperature–pressure grid, following the algorithm of \cite{1967ApJ...149..169M}, which is based on the Saha equation. This EOS accounts for the effects of both first and second atomic ionization for all elements up to $Z = 92$, as well as the formation of hydrogen molecules. The EOS results are stored in lookup tables, enabling efficient conversion between thermodynamic quantities.  For more details check \citet{2015AnGeo..33..703V} and
\cite{2024SoPh..299...23M}.

\subsection{Radiative terms}\label{subsec:rad}

The radiative loss term in the optically thick plasma of the sub-photospheric regions, photosphere, and low chromosphere (\qrad) is calculated by solving the non-gray radiative transfer (RT) equation under the assumption of LTE, using a domain decomposition following \cite{2005A&A...429..335V}. The wavelength dependence of emission and absorption coefficients is discretized through the opacity binning method \citep{1982A&A...107....1N}. For the simulations discussed in this paper, a single bin was used, corresponding to the gray approximation. Angle integration is performed using quadrature with three rays per octant. The RT module's formal solver is based on the short-characteristics method by \cite{1987JQSRT..38..325O}. A detailed description is provided in \cite{2024SoPh..299...23M}.

\begin{figure}
    \centering
    \includegraphics[height=0.2\textheight]{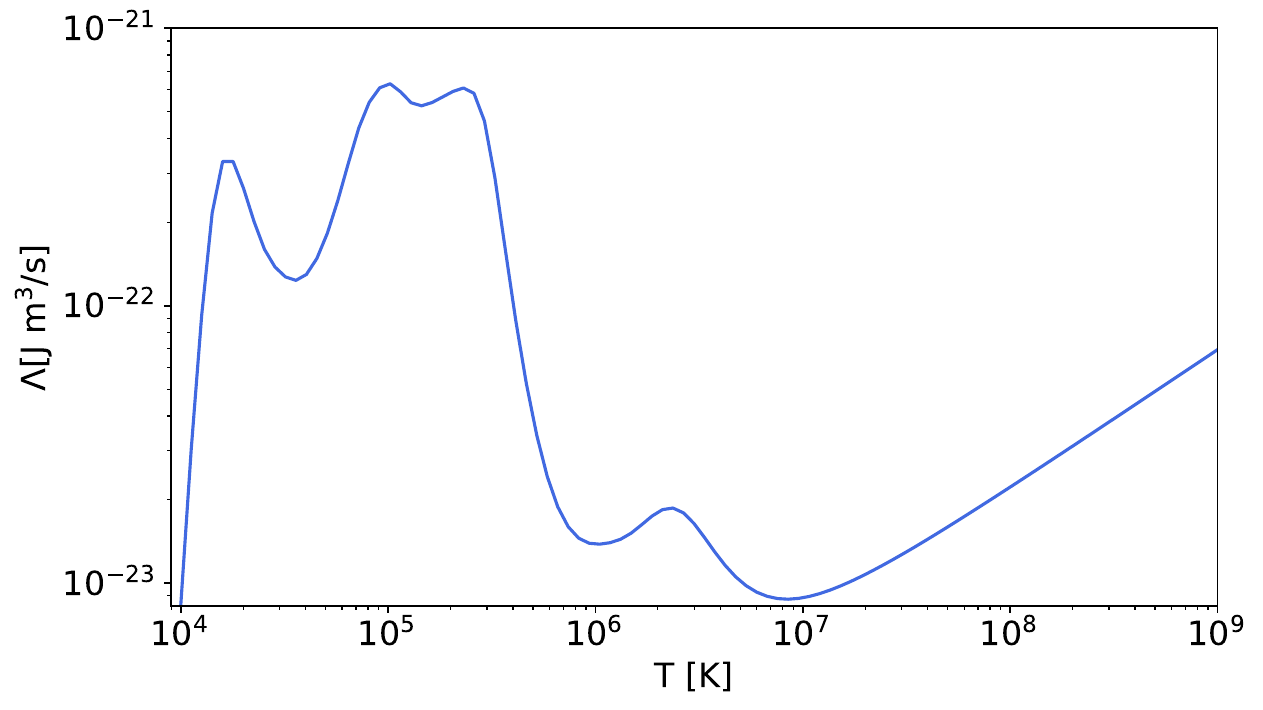}
    \caption{Density-averaged radiative loss rate $\Lambda$(T) from the CHIANTI 10.1 database with photospheric abundances.}
    \label{fig:chianti}
\end{figure}

While the LTE assumption limits the accuracy of the RT module in regions above the photosphere, we turn off this term in regions with temperatures higher than 20,000 [K], for which we use an approximation for radiative losses in the optically thin plasma of the form
\begin{equation}
\qlossEQ = -n_{e} n_{H}\Lambda(T),    
\end{equation}
where the electron/hydrogen number densities $n_e/n_H$ are given by 
\begin{eqnarray}
    n_e = \frac{\rho}{m_p} \frac{1+X}{2} \, ,
    n_H = \frac{\rho}{m_p} X \, ,
\end{eqnarray}
with $X = 0.7$ begin hydrogen abundance, and $m_p$ the proton mass. We use the tabulated loss function $\Lambda(T)$ from the CHIANTI 10.1 database \citep{2023ApJS..268...52D} with photospheric abundances. Since the loss rate function in the database is provided for a specific particle density, we average the loss function over the range of $10^{8}$ to $10^{12}$ particles per cm$^3$. The resulting loss function is illustrated in Fig. \ref{fig:chianti}. 

\subsection{Numerical resistivity and viscosity \label{subsec:numerical_resistivity_viscosity} }

To estimate the numerical resistivity and viscosity of the code, we performed a series of experiments in which we introduced additional terms into the source vector to account for both effects. The modified source vector is given by:
\begin{eqnarray}
\mathbf{S} = 
\begin{bmatrix}
0 \\
\rho \mathbf{g} + \nabla \cdot \mathbf{\Pi} \\
-\nabla \cdot \mathbf{q} + \qradEQ + \qlossEQ + \nabla\cdot \left(\mathbf{\Pi}\cdot \mathbf{v} \right)  - \nabla \cdot \left( \eta \mathbf{J} \times \mathbf{B} \right) \\
-\eta \mu_0 \mathbf{J}
\end{bmatrix} \, ,   
\end{eqnarray}
where $\eta$ is the resistivity and  $\mathbf{\Pi}$ is the viscous stress tensor given by
\begin{equation}
  \mathbf{\Pi} = \rho\nu \left( \nabla\mathbf{v} + \nabla\mathbf{v}^{T} - \frac{2}{3}\mathbf{I}\nabla \cdot \mathbf{v} \right)  \, ,
\end{equation}
with $\nu$ representing the kinematic shear viscosity. 

\begin{figure}
    \centering
    \includegraphics[height=0.2\textheight]{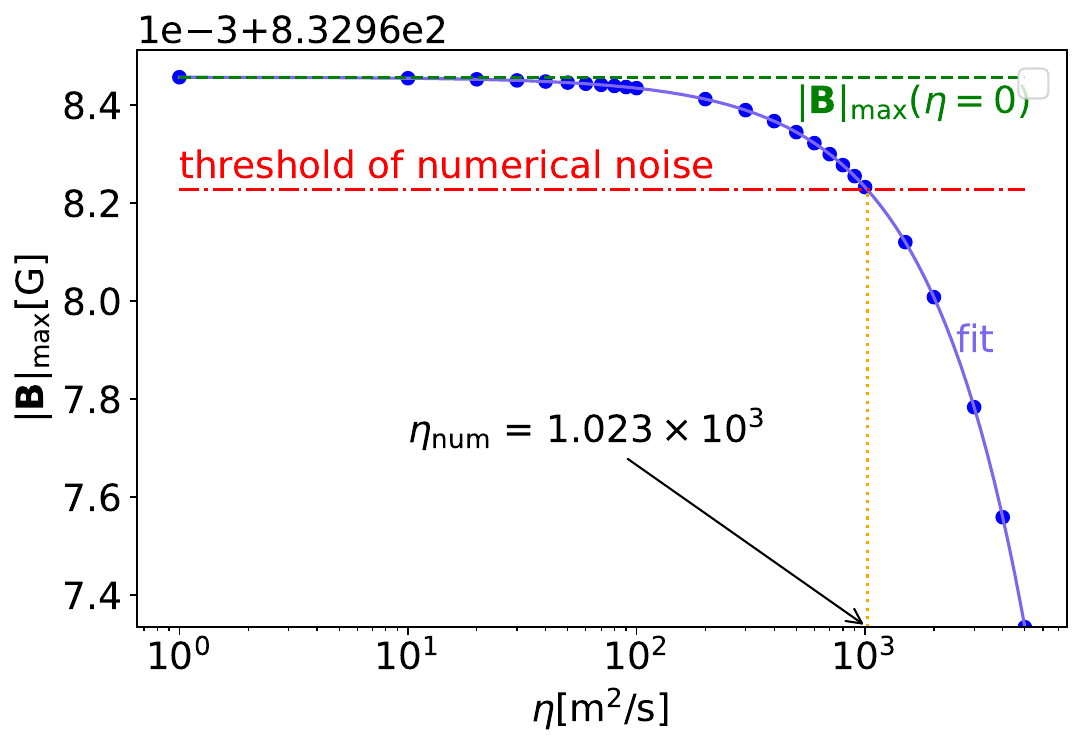}
   \includegraphics[height=0.2\textheight]{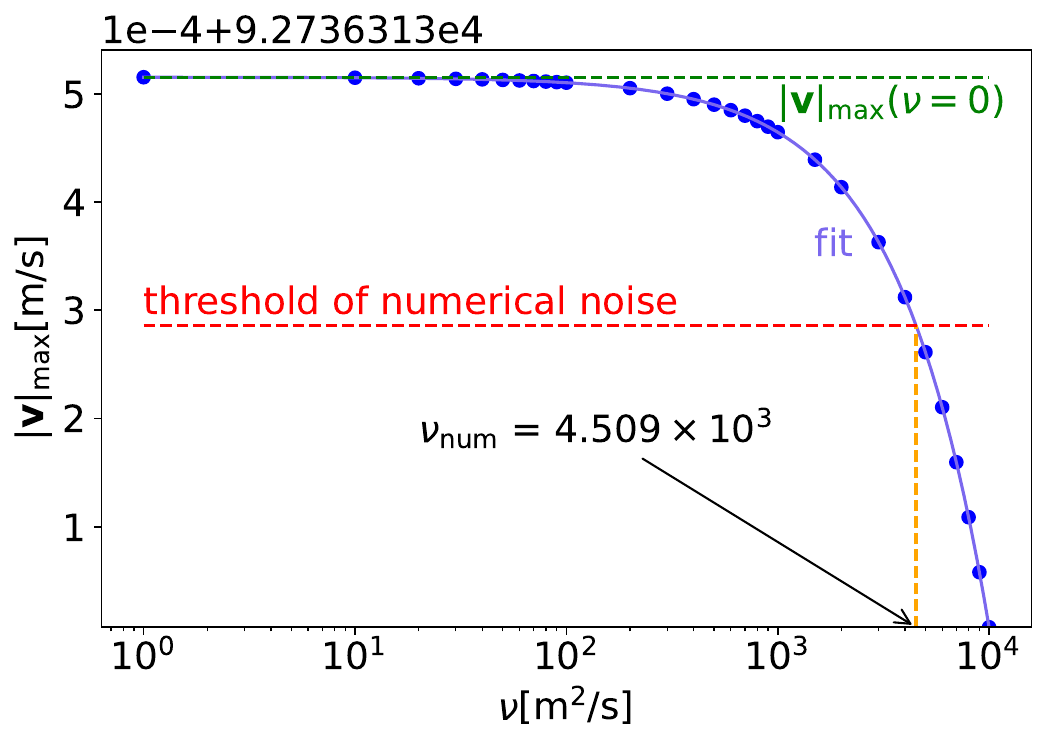}
    \caption{Graphical estimation of numerical resistivity (top panel) and numerical viscosity (bottom panel). Each dot represents the maximum magnetic field ($B_\text{max}$) or velocity ($V_\text{max}$) obtained from a simulation with a specific explicit value of $\eta$ or $\nu$. The blue solid line is a fit to these data points. Two horizontal lines are shown: one corresponds to the $B_\text{max}/V_\text{max}$ value for the case with $\eta=0$ and $\nu = 0 $, and the other marks the threshold of numerical noise. The intersection of this threshold with the fitted curve defines the estimated numerical resistivity or viscosity, indicated by the vertical line.}
    \label{fig:numerical_eta_nu}
\end{figure}

To evaluate the numerical resistivity in our experiments, we evolved a series of magneto-convection simulations using an arcade-like magnetic field configuration, varying the explicit resistivity. The specific details of the simulation setup will be presented in the following section. Starting from a baseline snapshot in the steady-state regime, we performed multiple simulations with different resistivity values, each evolved for one solar second. At the end of this period, we measured the maximum amplitude of the magnetic field. The top panel of Figure \ref{fig:numerical_eta_nu} shows the maximum value of the magnetic field amplitude decreasing as the resistivity value grows. 

The numerical resistivity is estimated by locating the value of $\eta$ for which the change in the maximum magnetic field amplitude is above the numerical noise threshold, such as that arising from round-off errors.
We determined the numerical noise threshold as follows. For each simulation, we computed the average of the maximum magnetic field amplitude, denoted as ${|\overline{\bf B|_{\rm max}}}_i$, along with its standard deviation $\sigma_i$. The ratio $\sigma_i/{|\overline{\bf B|_{\rm max}}}_i$  provides an estimate of the numerical variations that arise from round-off errors. We then identified the maximum value of this ratio in all simulations. Finally, we define the numerical noise threshold as the difference between the maximum magnetic field amplitude obtained from the zero-resistivity experiment and the largest numerical variation estimated among the simulations. This is 
\begin{equation}
\text{threshold of numerical noise} = |{\bf B}|_{\rm max}(\eta = 0)  - \text{max}\left(\frac{\sigma_i}{ {|\overline{\bf B|_{\rm max}}}_i }\right) \, .   
\end{equation}
The top panel of Figure \ref{fig:numerical_eta_nu} shows that the intersection of the fitting function with the noise threshold (horizontal red dashed line) is when $\eta = \eta_{\rm{num}} = 1.023\times 10^3$m$^2$/s. 
To estimate the numerical viscosity, we followed the same procedure but varied the values of $\nu$ and calculated the maximum speed instead of the maximum magnetic field amplitude. In this case, the numerical noise is
\begin{equation}
\text{threshold of numerical noise} = |{\bf v}|_{\rm max}(\nu = 0)  - \text{max}\left(\frac{\sigma_i}{ {|\overline{\bf v|_{\rm max}}}_i }\right) \, .   
\end{equation}
where $\sigma_i$ corresponds now to the standard deviation of the maximum speed in each simulation. The corresponding plot is presented in the bottom panel of Figure \ref{fig:numerical_eta_nu}. In such a case, we find that the numerical viscosity is $\nu_{\rm{num}} = 4.509\times 10 ^3$ m$^2$/s. 

To estimate the numerical resistivity, we focused on variations in the magnetic field amplitude, since resistivity appears in the induction equation and directly influences the magnetic field. While other physical quantities are also affected, the most significant variations are expected in the magnetic field. Following the same logic, we used the maximum velocity variations to estimate the numerical viscosity. Because viscosity acts as a dissipative term in the momentum equation, it primarily affects the momentum and velocity more rapidly than other variables.

It is important to note that the procedure used to estimate the numerical resistivity and viscosity provides only a simple approximation. The obtained values give an order-of-magnitude estimate, but in theory, they should not be constant. Moreover, the values we obtained may be valid only for our experiments, as they depend on the resolution, the Riemann solver, and the flux limiters employed. To obtain more accurate values, more specialized experiments should be carried out, for instance, as done by \cite{2017ApJS..230...18R}.

\section{Coronal simulations \label{sec:Simulations} }

\subsection{Initial set-up \label{subsec:ini_SetUp}}

\begin{figure}
    \centering
    \includegraphics[height=0.18\textheight]{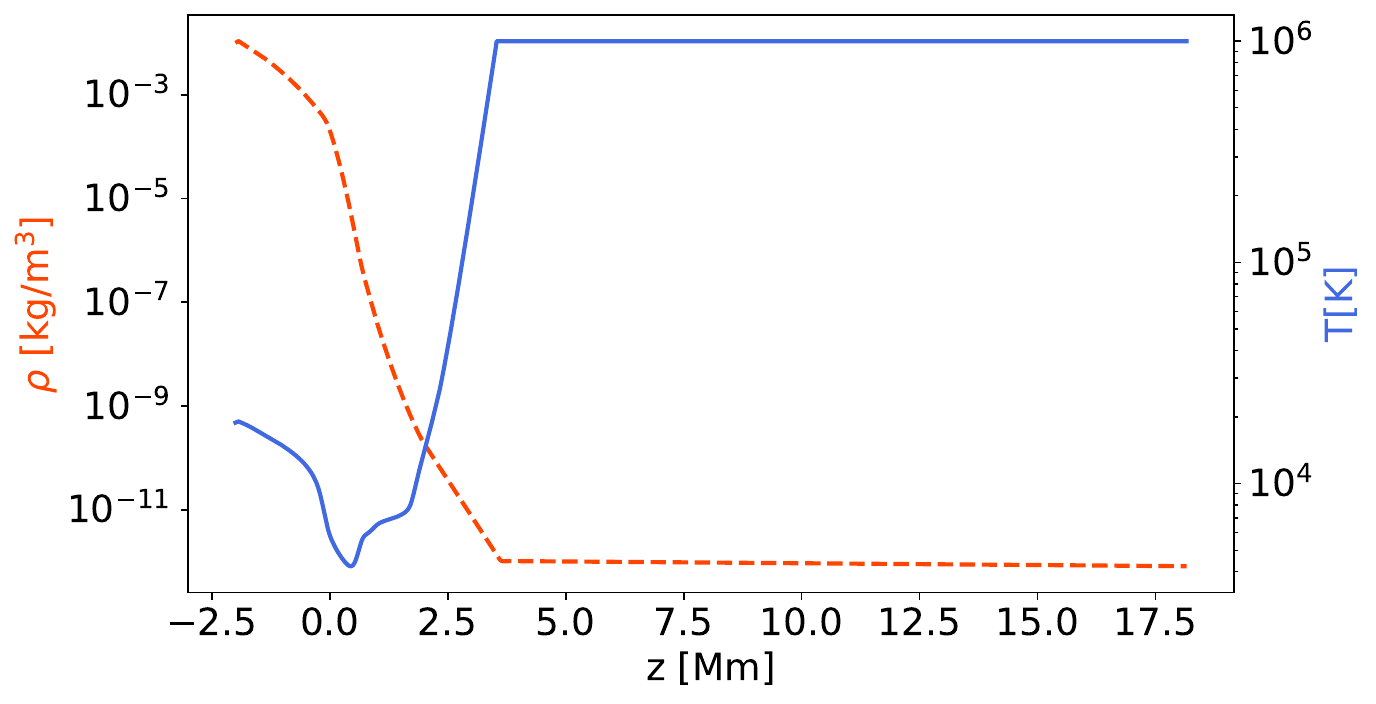}
    \caption{Initial profiles of density and temperature as functions of height.}
    \label{fig:ini_profile}
\end{figure}

The initial atmospheric structure for the simulations is constructed by combining several solar models. At heights $[-2,0]$ Mm, with $z=0$ defined as the height where the continuum optical depth at 500 nm equals 1, we employed the solar convection zone model from \cite{1974SoPh...34..277S}. This setup is connected to the VAL-C model \citep{Vernazza_temperature_profile} used in the height range [0, 2] Mm and is smoothly extended to $z = 3$ Mm where the temperature stabilizes at 1MK from [3, 18.16] Mm. The transition between models is achieved by imposing the hydrostatic equilibrium and ensuring consistency with the EOS. Figure \ref{fig:ini_profile} displays the initial density and temperature vertical profiles. 

In all simulations, the computational domain spans 46.08 $\times$ 20.16 Mm$^2$, extending from 2 Mm below the solar surface to 18.16 Mm above. The simulations use 2304 $\times$ 1440 points, providing a resolution of 20 km horizontally and 14 km vertically. Random noise was introduced to the internal energy to initiate the instability.

Horizontal boundary conditions (BC) are periodic, while the bottom boundary is open to allow mass flow, with automated control applied to regulate the fluctuation of the mass and entropy, as done in \cite{2018A&A...618A..87K}. The upper boundary has symmetric (zero gradient) conditions imposed on the internal energy, density, $v_x$, $v_y$, and $B_z$ and antisymmetric conditions on $B_x$ and $B_y$. In all simulations, the temperature at the top boundary is kept to its initial value. For the vertical velocity $v_z$, we use an outflow condition and it is set to zero when it is negative, thus preventing inflows, this is
\begin{align}    
    &{v_z}_{(N)} = \text{max}( {v_z}_{(N-2)} , 0)\nonumber \\
    &{v_z}_{(N-1)} = \text{max}( {v_z}_{(N-2)}, 0) \, ,
\end{align}
where ${v_z}_{(N)}$ and ${v_z}_{(N-1)}$ are the top boundary points.
\begin{figure}
    \centering
    \includegraphics[height=0.18\textheight]{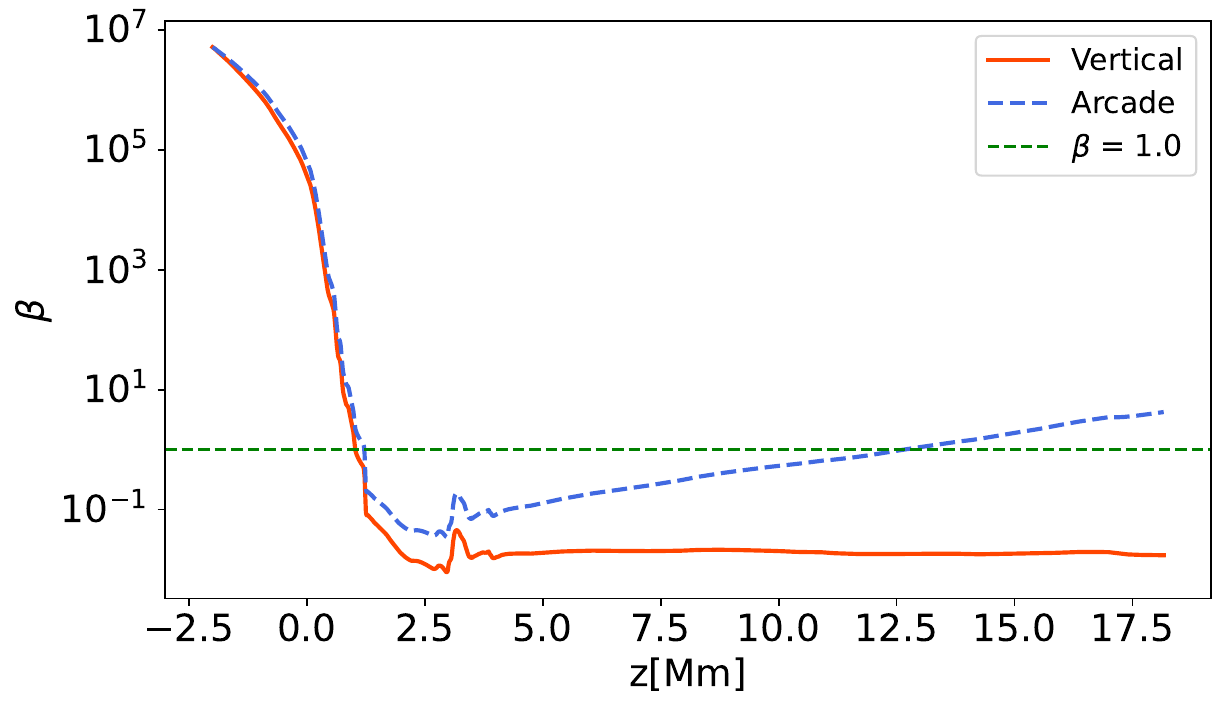}
    \caption{Initial plasma-beta profiles as a function of height, evaluated at the center of the domain in the $x-$direction, for both the vertical and arcade magnetic field configurations after field implantation.}
    \label{fig:ini_plasmabeta}
\end{figure}
\begin{figure}
    \centering
    \includegraphics[width=0.5\textwidth]{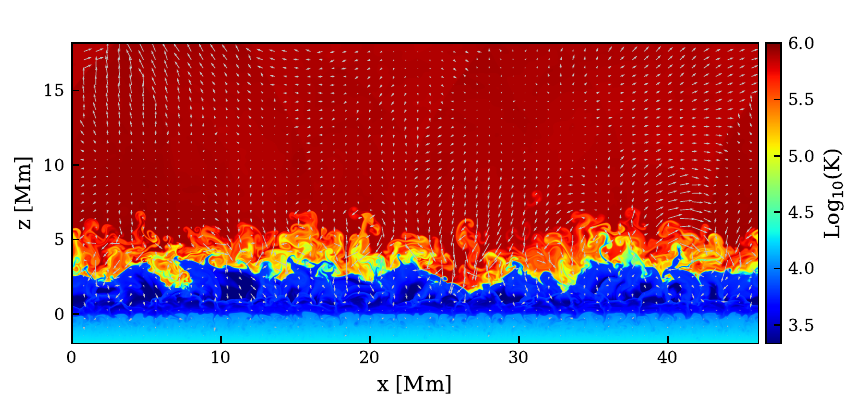}
    \caption{Colormap of the temperature from the baseline hydrodynamic simulation at $t=1000$ s. Arrows indicate the velocity field, with lengths proportional to the local flow speed. The full temporal evolution is available as an online movie (see links in Table~\ref{tab:description}).}
    \label{fig:Arc10_hydro}
\end{figure}

\begin{figure}
    \centering
    \includegraphics[height=0.18\textheight]{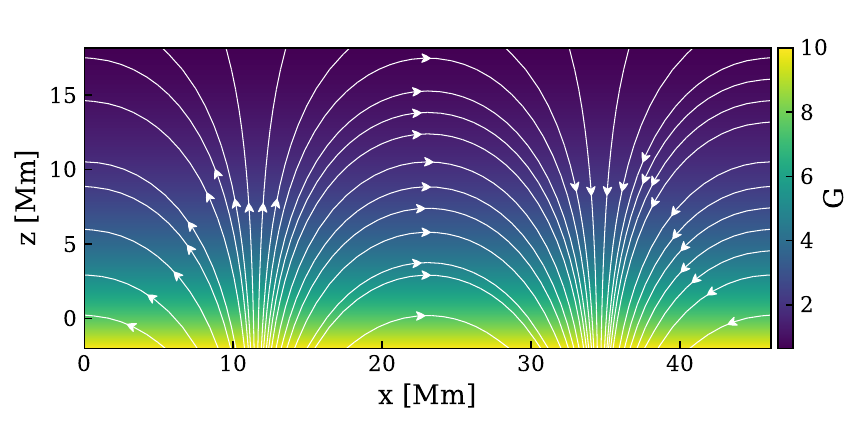}
    \caption{Colormap of the magnetic field strength in gauss and field line configuration of the arcade-like model.}
    \label{fig:arcade_alone}
\end{figure}

We first develop the convection instability in a simulation without a magnetic field, with \qtc = 0 and \qloss = 0. Once convection is fully developed and the system stabilizes, we extract a snapshot at $t=1000$ s, which we use as the base state for all subsequent simulations. Figure \ref{fig:Arc10_hydro} shows a logarithmic colormap of the temperature for this snapshot, overlaid with velocity field arrows, whose lengths are proportional to the local flow speed. At sub-surface depths, the plot reveals well-developed convective granules. However, in the overlying layers, the upward flows generated by the convective motions disrupt the coronal plasma, resulting in disorganized and unstructured mixing. This behavior is a consequence of the absence of a magnetic field, which in reality plays a crucial role in regulating and structuring coronal dynamics. While this scenario does not occur in nature, it serves as an intermediate step in the development of our models. 

After 1000s add either of the two different magnetic fields to the base hydrodynamic convective structure: the first one is a 10 G vertical magnetic field, the second one is a potential arcade-like magnetic field given by 
\begin{eqnarray}
& & B_x =  B_0 \cos(k x) \exp(-k (z - z_\text{0}))  \, , \\
& & B_z = -B_0 \sin(k x) \exp(-k (z - z_\text{0}))   \, ,  \label{eq:arcade}
\end{eqnarray}
where $B_0 = 10$ G, $z_0 =-2$ Mm,  $k = 2 \pi/L_x$ and $L_x = 46.08$ Mm is the horizontal size of the domain (figure \ref{fig:arcade_alone}). The magnetic field is either constant or potential and therefore is initially consistent with the equations because it is force-free. Then, convection smoothly moves the magnetic field to make it consistent with the flow structures.

Figure \ref{fig:ini_plasmabeta} presents one-dimensional vertical profiles of the plasma-beta for each implanted magnetic field configuration, evaluated at the center of the domain along the $x-$direction. The horizontal line indicates the location where plasma-beta equals one, marking the transition between pressure-dominated and magnetically dominated regions. Both configurations become magnetically dominated at approximately the same height ($z\approx 1$ Mm), though the vertical field case exhibits a lower plasma-beta compared to the arcade configuration. In the arcade case, plasma-beta rises again, indicating a return to pressure domination around $z\approx12$ Mm.

\subsection{Results\label{subsec:results}}

We performed a total of 4 simulations evolved for one hour of solar time, including a simulation featuring the arcade-like magnetic field, with Spitzer's thermal conduction model. We use the minmod slope-limiter, the HLLC Riemann solver, and the SSP-RK3 time integration method. Table \ref{tab:description} summarizes the main characteristics of each simulation and includes hyperlinks to animations of the density and temperature. The \textit{Field} column indicates the magnetic field configuration, and \textit{TC} refers to the thermal conduction model applied (either Braginskii, B; or Spitzer, S).
\begin{table}
\renewcommand{\arraystretch}{1.2}
    \caption{Summary of simulation parameters, magnetic field configurations, thermal conductivity (TC) models (Braginskii, B; or Spitzer, S), initial and final times, and corresponding animation links. }
    \centering
 \begin{tabular}{|c|c|c|c|c|c|}
 \hline
 \#  & Field                      & TC                       & $t_\text{ini}[s]$ & $t_\text{fin}[s]$ &   Animation links            \\ \hline
  1  & none                       & none                           & 0                     & 1000 &                  T (link 1), $\rho$ (link 2)  \\ \hline
  2  & vertical                   & \multirow{2}{*}{B}  & 1000                  & 4600 &                  T (link 3), $\rho$ (link 4)  \\ \cline{1-2} \cline{4-6} 
  3  & \multirow{2}{*}{arcade}    &                                & 1000                  & 4600 &                  T (link 5), $\rho$ (link 6)  \\ \cline{1-1} \cline{3-6} 
  4  &                            & S                      & 1000                  & 4600 &                  T (link 7), $\rho$ (link 8)  \\ \cline{1-1}  \hline
  \end{tabular} 

    \label{tab:description}
\end{table}

\subsubsection{Overview of the simulations}
\begin{figure}
    \centering
    \includegraphics[width=\linewidth]{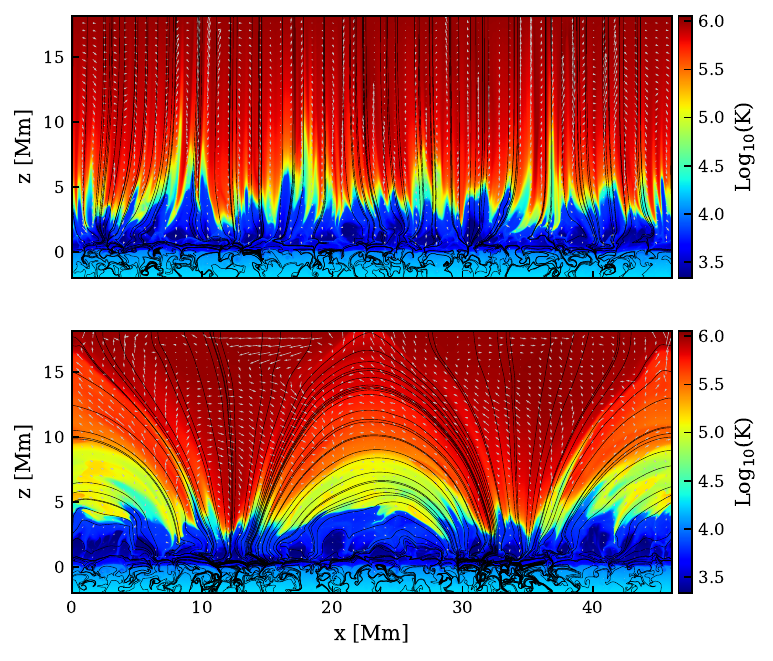}
    \caption{Colormap of the logarithmic temperature distribution after 1 hour of solar time. The top panel shows the simulation with a vertical magnetic field (Simulation 2), and the bottom panel shows the simulation with an arcade-like magnetic field (Simulation 3). The arrows indicate the velocity field, with lengths proportional to the local flow speed, black lines are magnetic field lines.} 
    \label{fig:te_colormap_Arc_V}
\end{figure}

Intermediate states from Simulations 2 and 3 are shown in Figure \ref{fig:te_colormap_Arc_V}, corresponding to the vertical and arcade-like magnetic field configurations, respectively. The snapshots were taken at 30 minutes of solar time—halfway through the total simulation duration. The full-time evolution for all simulations is available via the links provided in Table \ref{tab:description}. The colormap represents the logarithm of the temperature, with magnetic field lines overlaid using equally spaced seed points. Velocity vectors are also plotted, with arrow lengths proportional to the local flow speed.

A prominent feature in both cases is the preservation of the overall magnetic field topology above the surface. In Simulation 2, the field retains its predominantly vertical configuration, while in Simulation 3, the arcade-like structure remains largely intact. In both simulations, elongated structures form along the magnetic field lines.

The transition region behaves differently between the two configurations. In the arcade-like field simulation, it appears significantly broader beneath the loop system, where temperatures remain cooler than in areas with a predominantly vertical field.

The velocity field exhibits a clear trend with height: the flow speed generally increases with altitude. However, the flow does not strictly follow the direction of the magnetic field—particularly in the arcade-like configuration, where significant deviations are observed in the upper regions. These deviations occur mainly in areas where the plasma $\beta > 1$, as evident from the plasma $\beta$ profile shown in Fig.~\ref{fig:ini_plasmabeta}. In the regions where $\beta < 1$, the flow is more closely aligned with the magnetic field; nevertheless, the alignment is not exact, since the plasma $\beta$ is less than unity but not sufficiently small to enforce a strong coupling. In contrast, in the vertical field case, the inclination of the velocity vectors relative to the magnetic field lines is considerably smaller, indicating a more aligned flow, as expected.

Below $z = 0$ Mm, in the sub-convection region, both simulations exhibit similar thermal and dynamic behavior. The temperature profiles are comparable, and the granulation pattern—including cell size—is consistent across both runs. However, the magnetic field evolution differs: in the arcade-like field simulation, strong concentrations of magnetic field develop near the footpoints, forming localized, high-intensity braided structures.

\begin{figure*}
    \centering
    \includegraphics[width=0.45\linewidth]{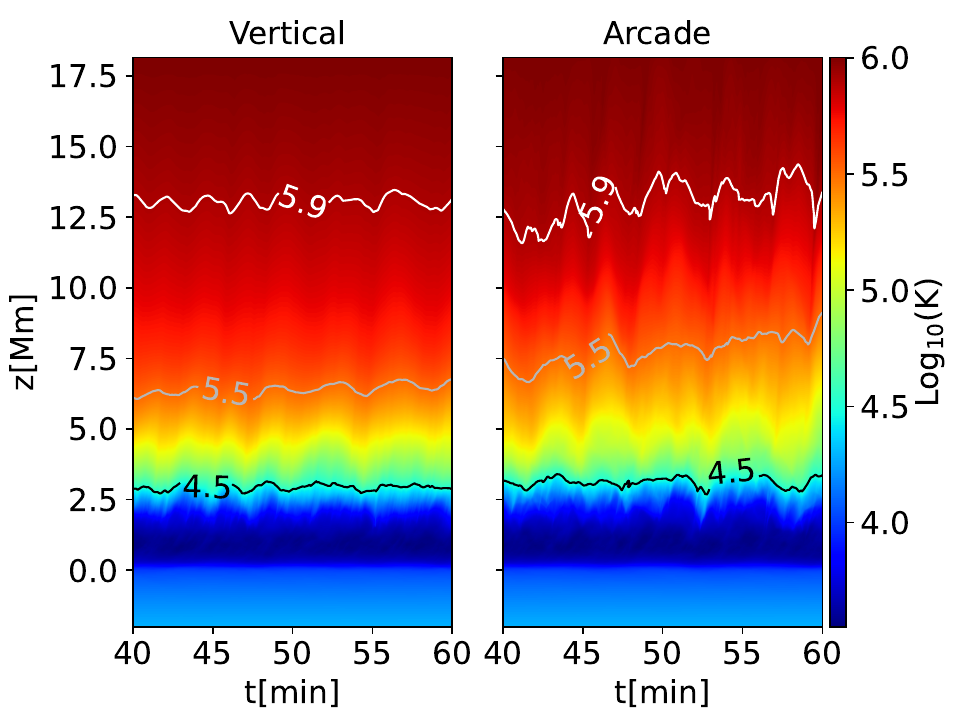}
    \includegraphics[width=0.42\linewidth]{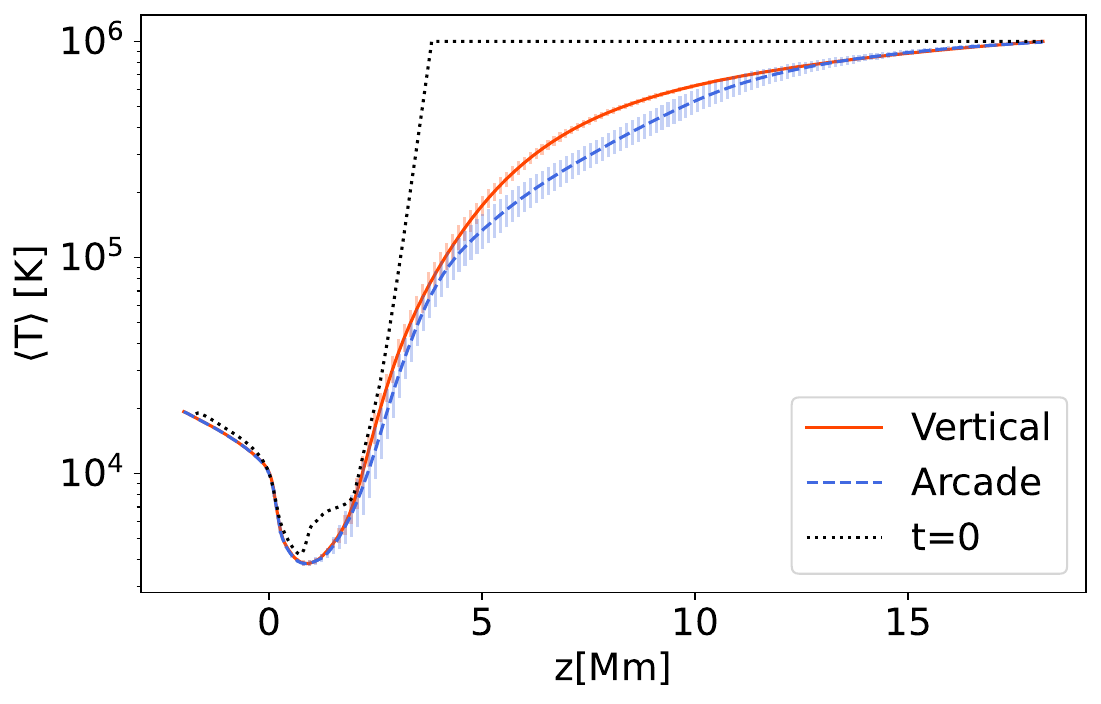}
    \caption{Horizontally averaged temperature profiles for the simulations shown in Fig.~\ref{fig:te_colormap_Arc_V} for the last 20 minutes. The left panel displays the horizontally averaged temperature as a function of height and time, with contour lines overlaid to highlight specific temperature values. The right panel shows the temperature averaged both horizontally and over time. The black dotted line indicates the initial profile.}
    \label{fig:average_te_Arc_vs_V}
\end{figure*}

For the statistical analysis, we compute the horizontal average of the temperature over the last 20 minutes of each simulation. The left panel of Figure \ref{fig:average_te_Arc_vs_V} shows the horizontally averaged temperature as a function of height and time. The colormap on the left corresponds to the vertical magnetic field case, while the one on the right corresponds to the arcade-like magnetic field case. Contour lines are overlaid to highlight three specific temperature levels:  $10^{4.5}$ K, $10^{5.5}$ K, and $10^{5.9}$ K. The first contour, $10^{4.5}$ K, is characteristic of chromospheric temperatures; the second contour, $10^{5.5}$ K, also shows little temporal variation and appears to maintain a height around $z = 6$ Mm in the vertical field case but about 1 Mm higher in the arcade-like configuration. This indicates that, beyond the transition region, the vertical field case maintains a hotter corona than the arcade-like case. The third contour, $10^{5.9}$ K, shows greater oscillations in the arcade-like case, consistent with Figure \ref{fig:te_colormap_Arc_V}, which shows that the coronal temperature depends in this case on the magnetic field inclination. 

 The right panel of Figure \ref{fig:average_te_Arc_vs_V} presents the temperature averaged both horizontally and over time. The black dotted line indicates the initial temperature profile. The most significant difference between the two cases occurs in the height range $z =$ [5, 10] Mm, where the vertical field case retains a much hotter corona. The maximum temperature difference between the two models reaches approximately 135 kK. Comparing the time-averaged profiles to the initial state clearly shows that neither of the two models achieves a restoration of the chromospheric heating.  Vertical error bars represent the standard deviation, which quantifies the variations of the horizontally averaged temperature over time. The variations are minimal in the solar interior ($z<0$ Mm), peak in the transition region, and decrease again towards the corona. The vertical field case exhibits smaller deviations overall, consistent with both the left panel and the colormaps of Figure \ref{fig:te_colormap_Arc_V}.  

\begin{figure}
    \centering
  \includegraphics[width=0.9\linewidth] {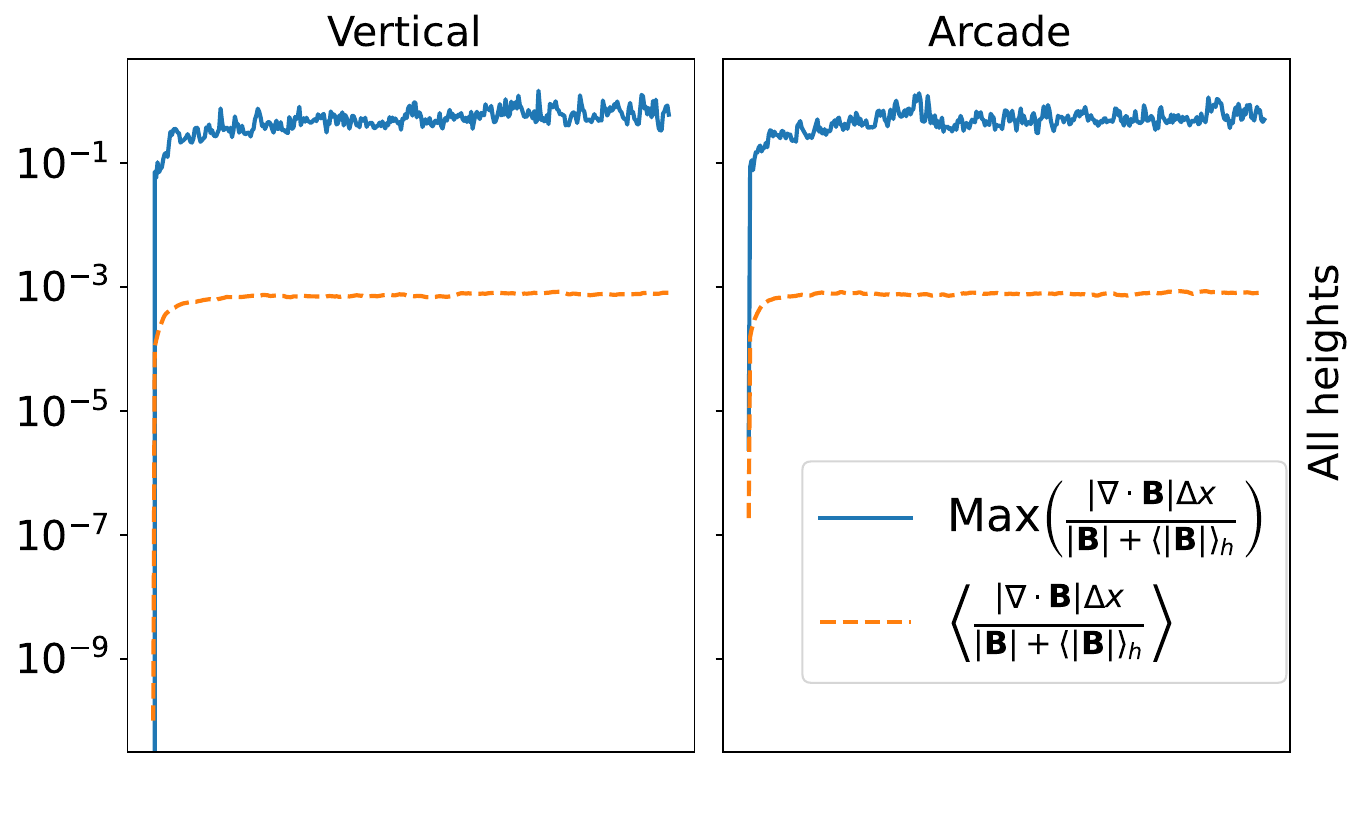}
  \includegraphics[width=0.9\linewidth]{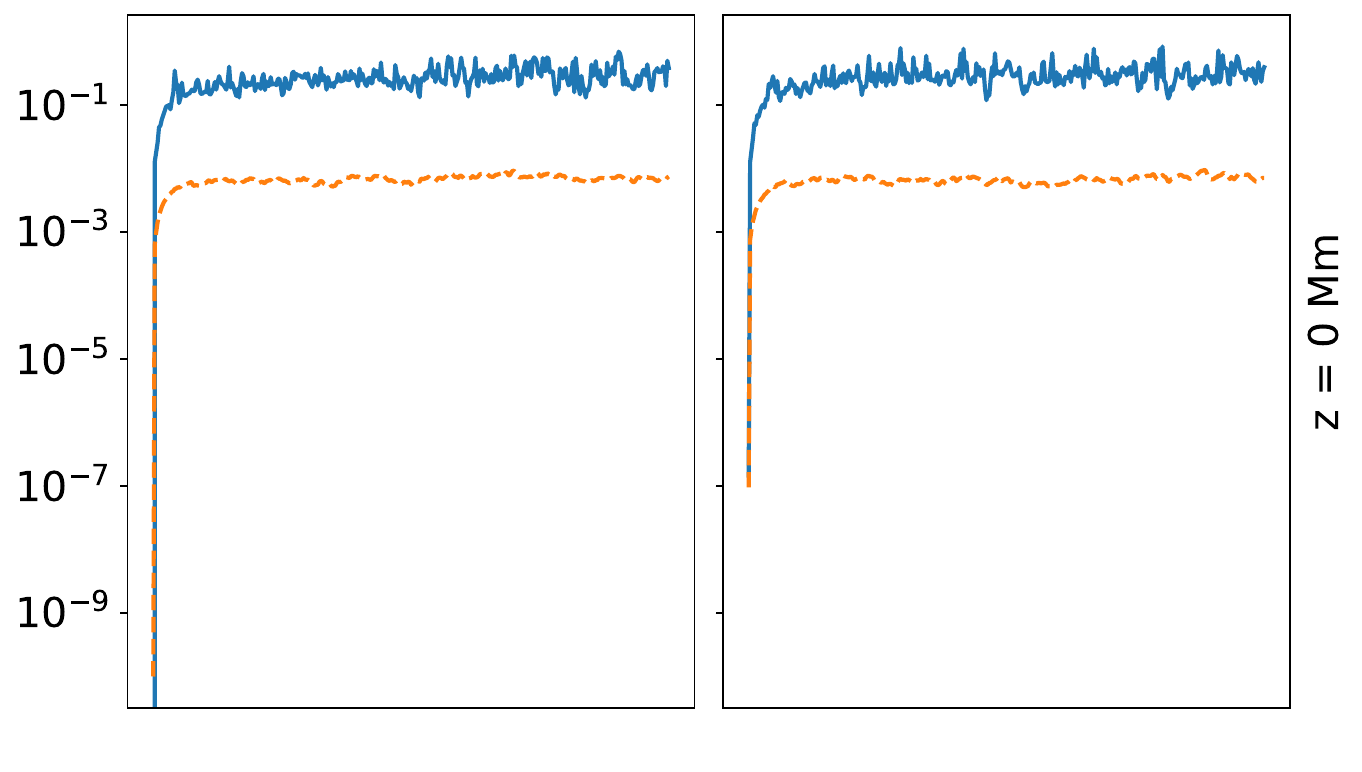 }
  \includegraphics[width=0.9\linewidth]{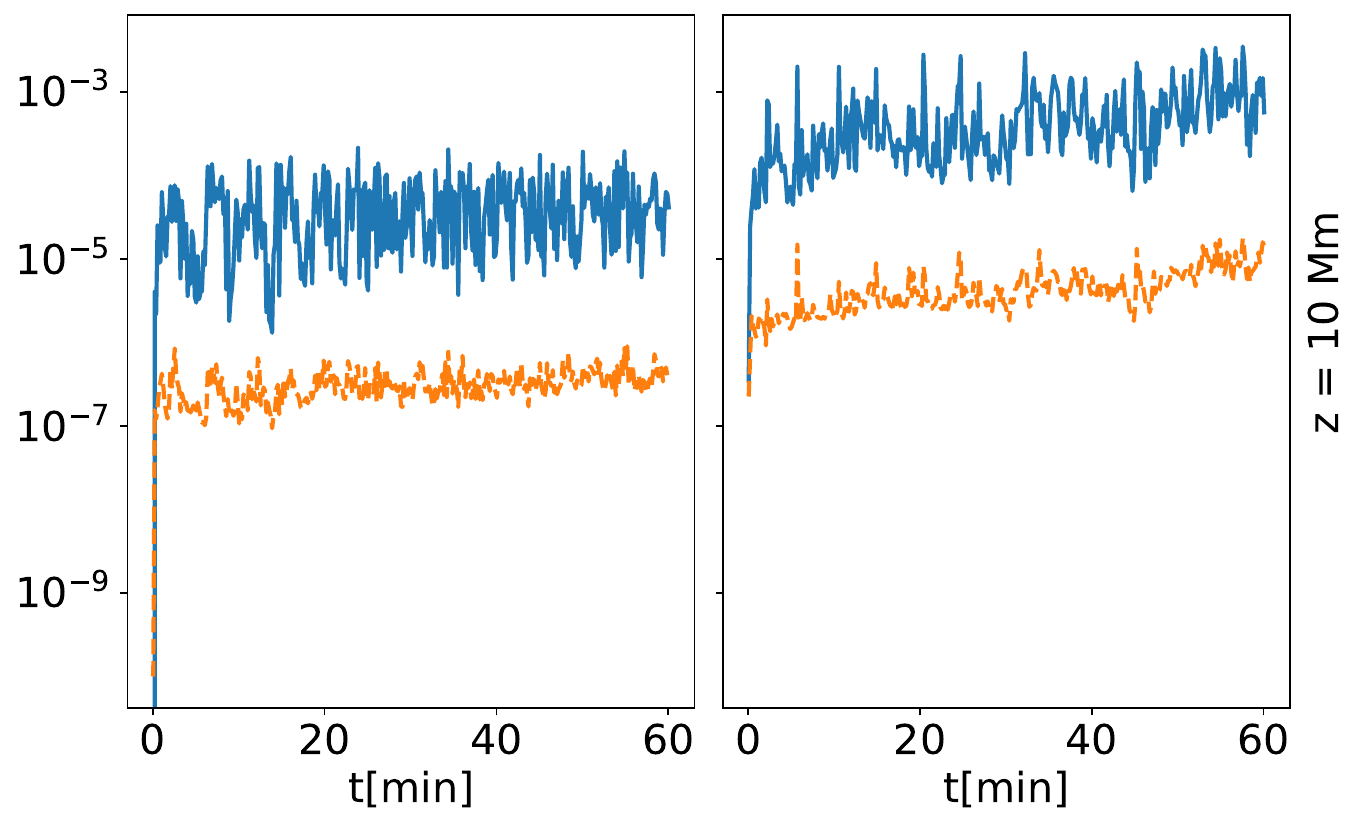 }
 \includegraphics[width=0.9\linewidth]{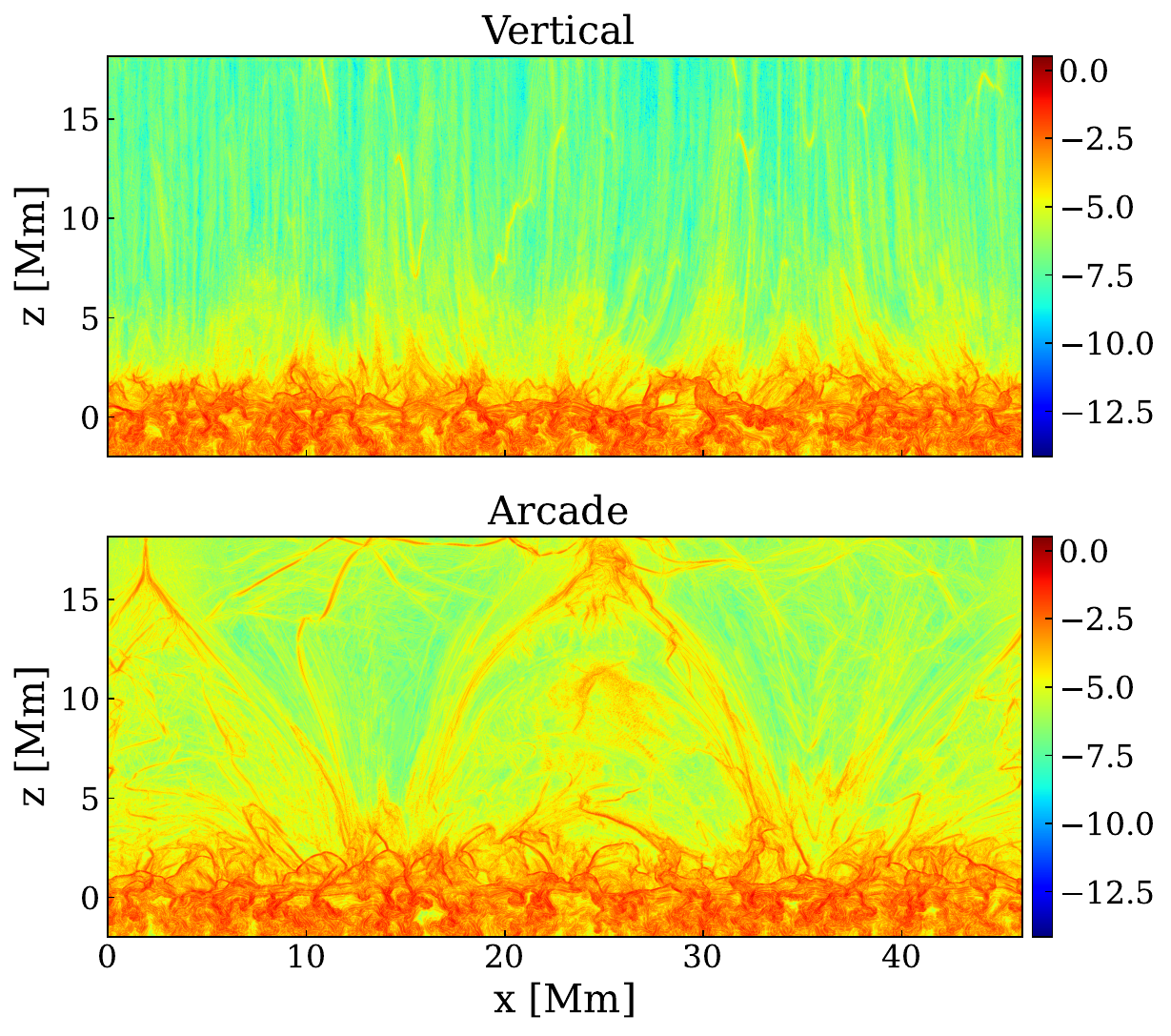 }
\caption{Dimensionless error associated with the divergence of the magnetic field, defined as $|\nabla\cdot{\bf B}|\Delta x/(|{\bf B}| + \left<|{\bf B}|\right>_h)$. The top three panels show the maximum and average values of this error as functions of time. The left panels correspond to the simulation with an initially vertical magnetic field, while the right panels show the results for the arcade magnetic field configuration. The first row presents values computed over the entire domain, while the second and third rows show values at heights of $z$=0 Mm and $z$=10Mm, respectively. The bottom colormaps display the error in logarithmic scale for both simulations for $t$=2800s.} \label{fig:divB}
\end{figure}

To illustrate the overall behavior of the magnetic field divergence, Figure \ref{fig:divB} shows the time evolution of the maximum and average values of the  dimensionless error associated to the divergence of the magnetic field defined as $|\nabla\cdot{\bf B}|\Delta x/(|{\bf B}| + \left<|{\bf B}|\right>_h)$. Here, the value of the average of the magnetic field is used to avoid indeterminations. The plot in the top corresponds to all the values in the domain, and the other two correspond to the heights $z$= 0 and $z$ = 10 Mm. The difference between the average and maximum errors spans approximately two orders of magnitude, both across the full domain and at fixed heights. Furthermore, the error is larger in the lower regions of the atmosphere and decreases in the upper layers, consistent with a stronger field in the near-surface convection. The bottom panels display the logarithmic colormap of the error for an intermediate snapshot at $t$=2800s, shown for both simulations. These results demonstrate that, on average, the constrained transport method \citep{CT_evans} is effective in maintaining the solenoidal condition of the magnetic field. However, they also indicate that the method does not completely eliminate local violations of the divergence-free constraint.

\subsubsection{Energy contributions}

To study the processes that most strongly influence the thermal structure of the atmosphere in our simulations, we now consider the evolution equation for the internal energy:
\begin{eqnarray}
& & \frac{\partial e_{\rm int}}{\partial t} = -\nabla\cdot\left(e_{\rm int} \bf{v}\right) -  p\nabla \cdot {\bf v}  - \nabla \cdot \bf{q} +  \qlossEQ + \qradEQ \nonumber \\ 
& & \hspace{1.3cm} + \nabla\cdot\left({\bf \Pi}_{\rm num} \cdot \bf{v}\right) + \eta_{\rm num}J^2 \, , \label{eq:evol_eint}
\end{eqnarray}
where $\nabla\cdot\left(e_{\rm int} \bf{v}\right)$ represents the internal energy flux, $p\nabla \cdot {\bf v}$ the pressure work, $\nabla \cdot \bf{q} $ the thermal conduction contribution, $\qlossEQ$ the optically thin radiative losses, $\qradEQ$ the optically thick radiative term. We have included the numerical resistive heating  $\eta_{\rm num}J^2$, and the numerical viscous dissipation term $\nabla\cdot\left({\bf \Pi}_{\rm num} \cdot \bf{v}\right)$, however both of them are implicit terms. We estimated the values of $\eta_{\rm num}$ and $\nu_{\rm num}$ back in Section \ref{subsec:numerical_resistivity_viscosity}. 

\begin{figure*}
    \centering
    \includegraphics[width=0.8\linewidth]{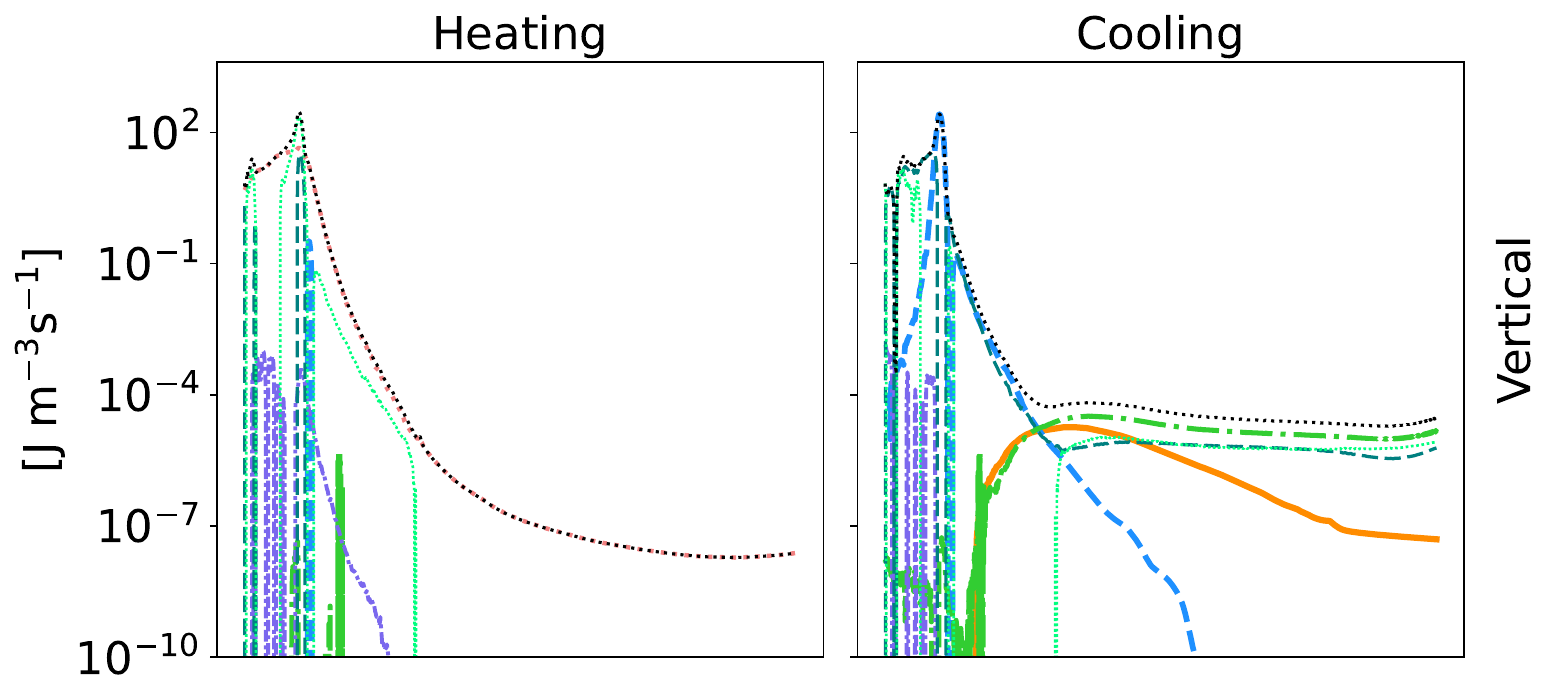}
    \includegraphics[width=0.8\linewidth]{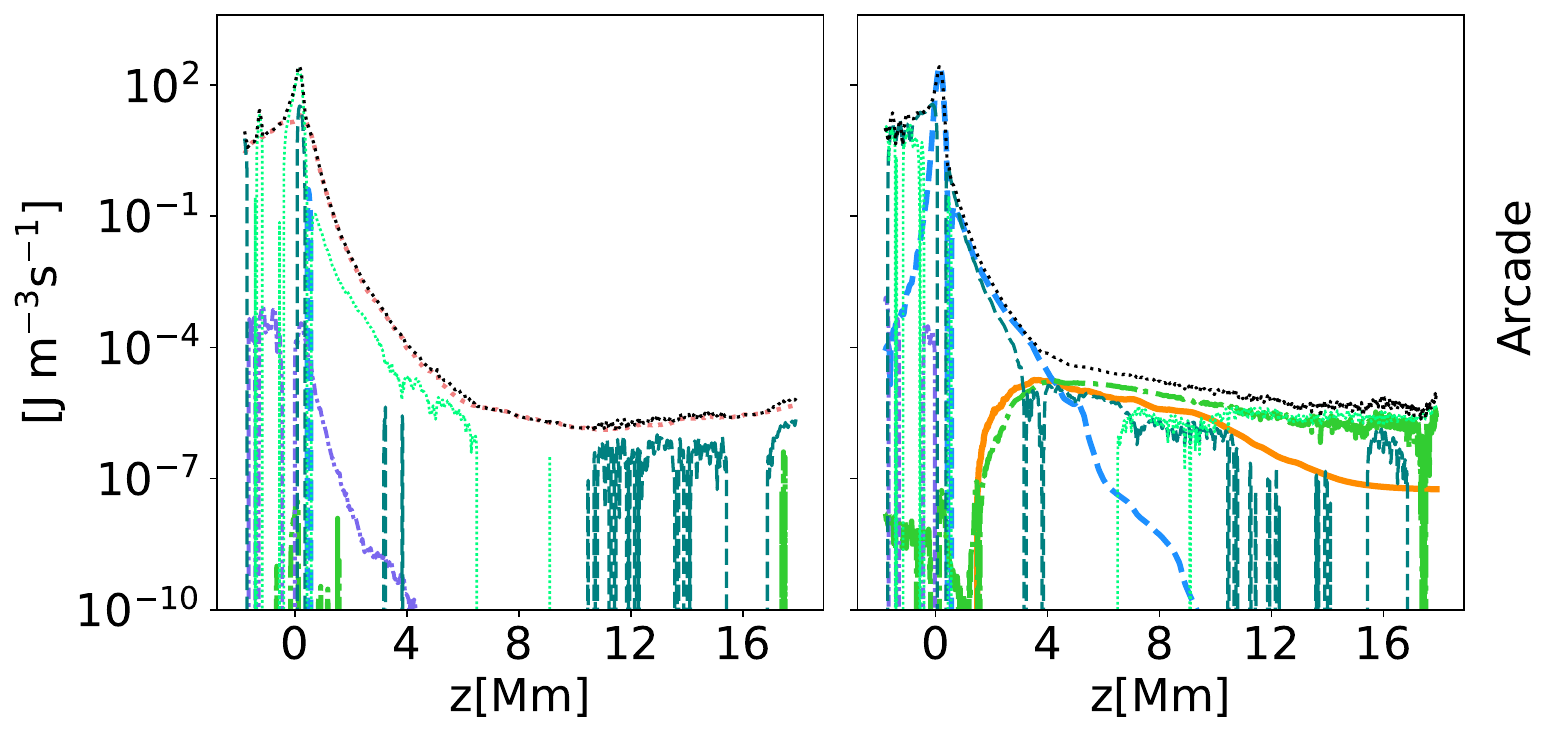}
    \includegraphics[width=0.6\linewidth]{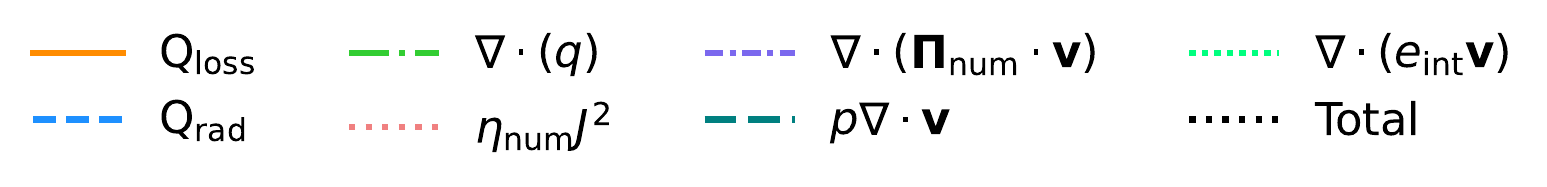}
    \caption{Horizontally and temporally averaged heating and cooling contributions to the internal energy equation as a function of height. The plotted terms are the radiative terms \qloss (orange), \qrad (blue), divergence of heat flux $(\nabla \cdot \bf{q})$ (green), joule heating from numerical resistivity ($\eta_{\rm num}J^2$, (red)), dissipation from numerical viscosity $(\nabla\cdot({\bf \Pi}_{\rm num} \cdot \bf{v}))$, (violet), adiabatic heating $(p\nabla \cdot {\bf v})$,  (navy blue), and flux of internal energy $(\nabla\cdot (e_{\rm int} \bf{v}))$, (light green).  Left panels show heating terms, and right panels show cooling terms. The top row corresponds to the simulation with a vertical magnetic field (Simulation 2), while the bottom row corresponds to the simulation with an arcade-like magnetic field (Simulation 3). The black dotted line in each panel represents the total contribution (sum of all terms). For clarity, values smaller than $1\times10^{-10}$ are omitted from the plots.}
    \label{fig:eint_eq}
\end{figure*}
\begin{figure*}
    \centering
    \includegraphics[width=0.35\linewidth]{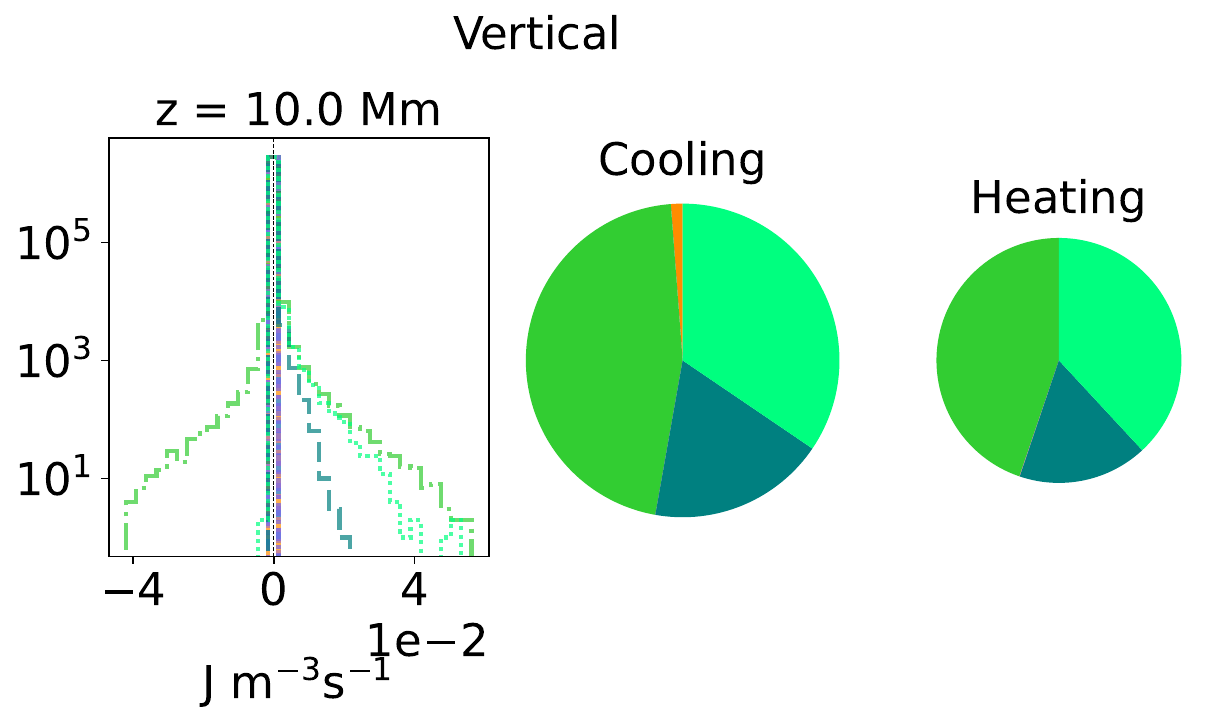}
    \includegraphics[width=0.35\linewidth]{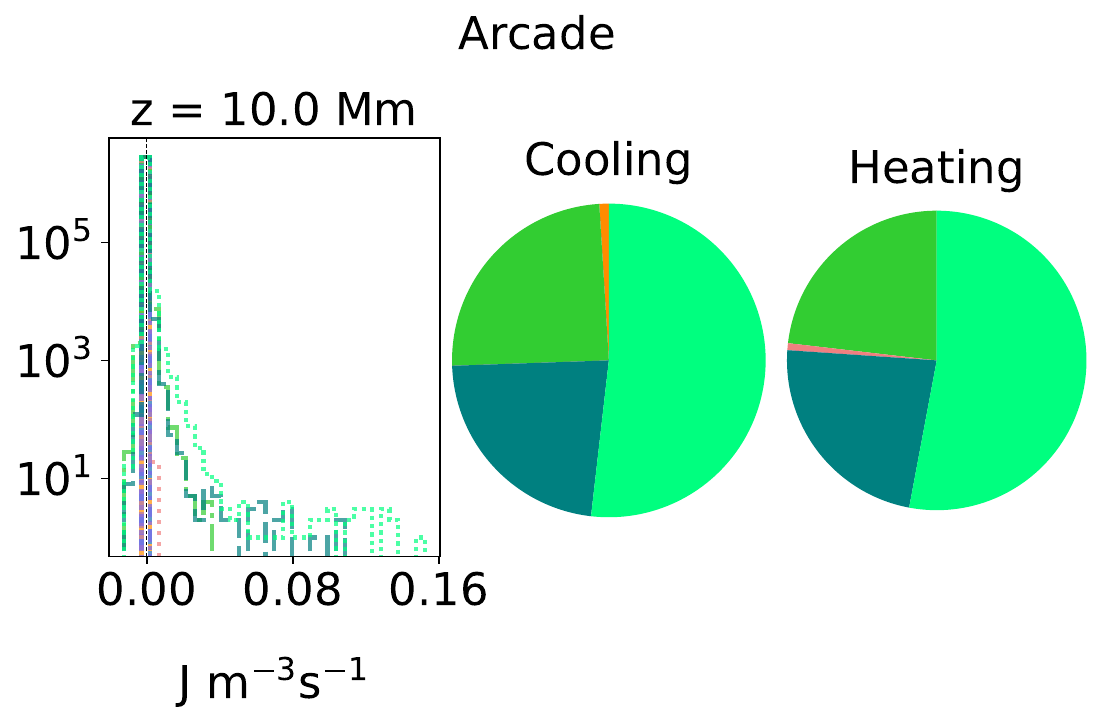}
    \includegraphics[width=0.35\linewidth]{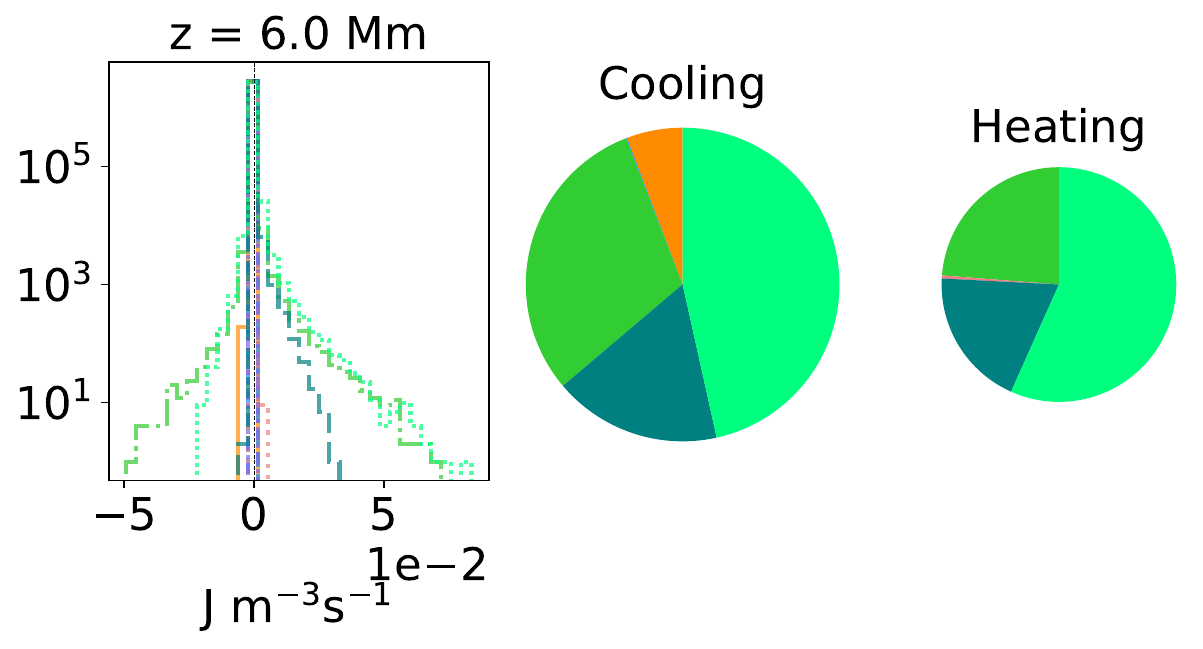}
    \includegraphics[width=0.35\linewidth]{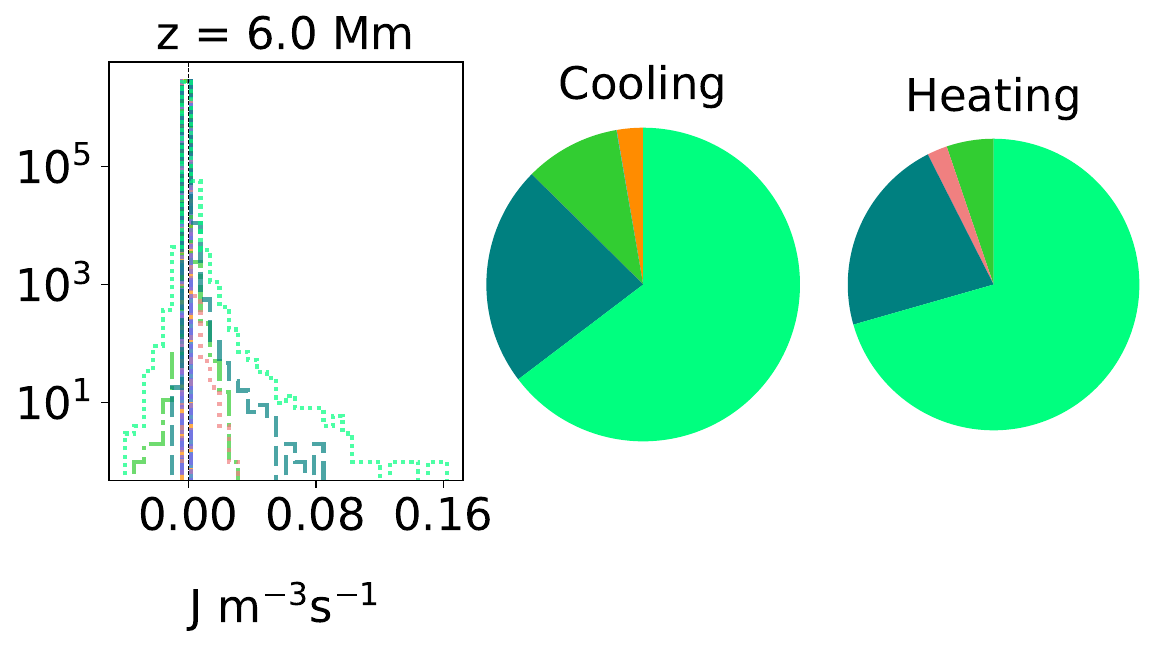}
    \includegraphics[width=0.35\linewidth]{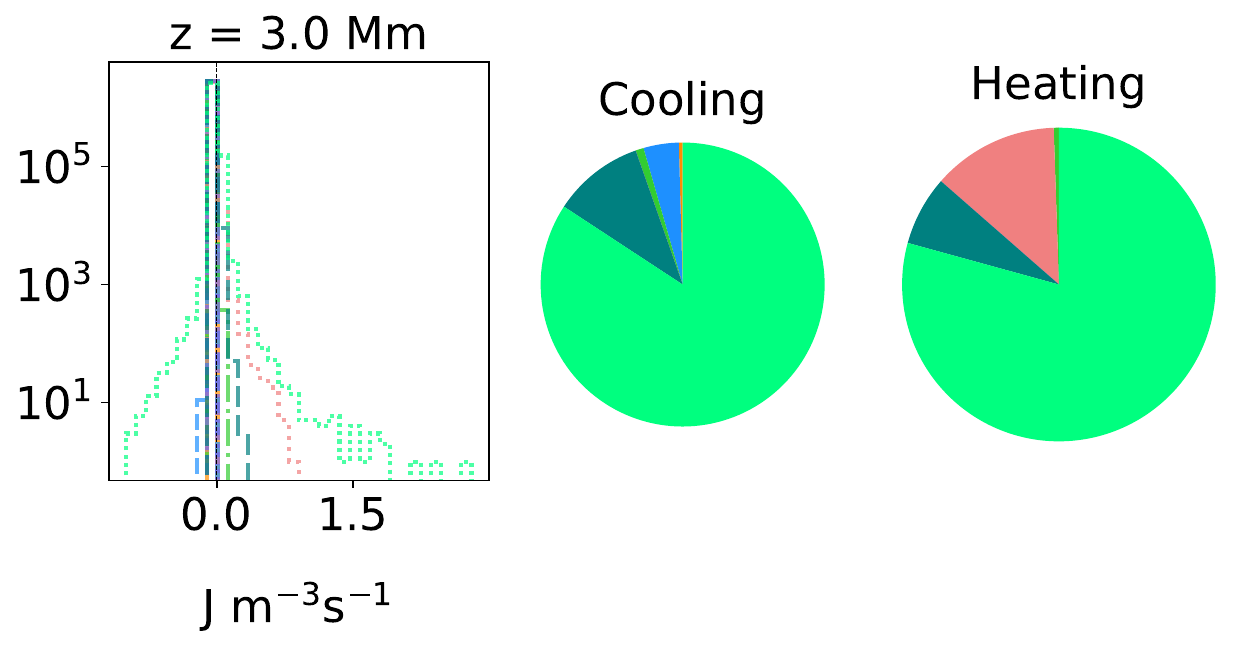}
    \includegraphics[width=0.35\linewidth]{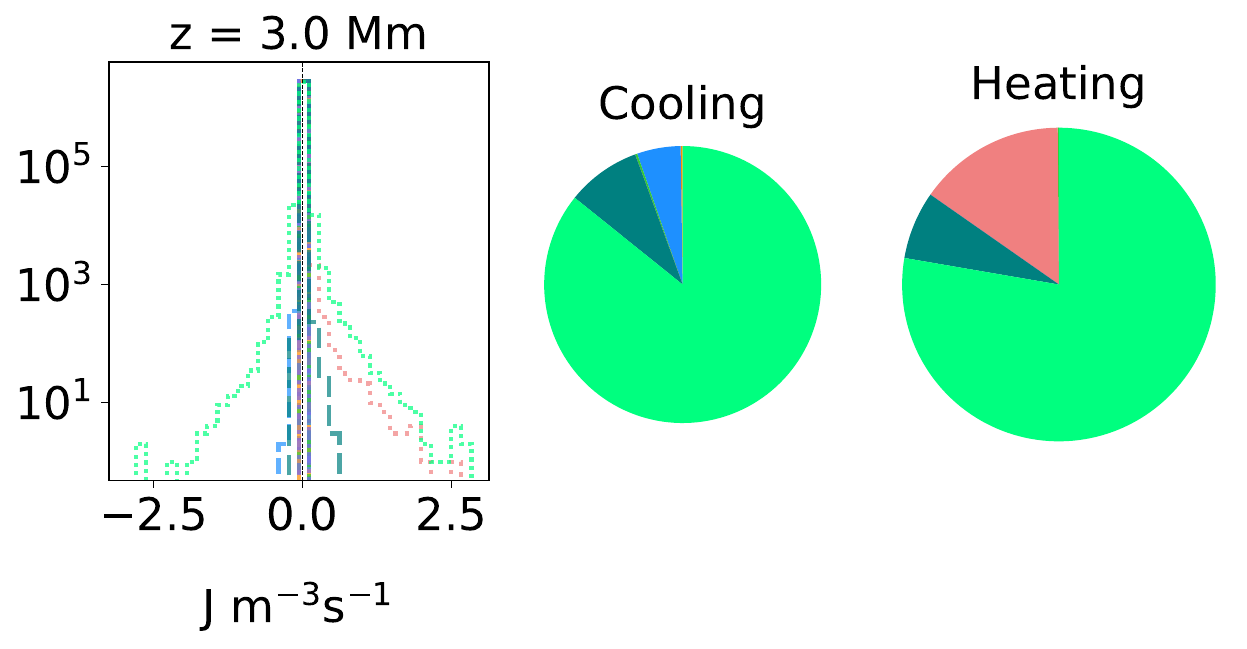}
    \includegraphics[width=0.35\linewidth]{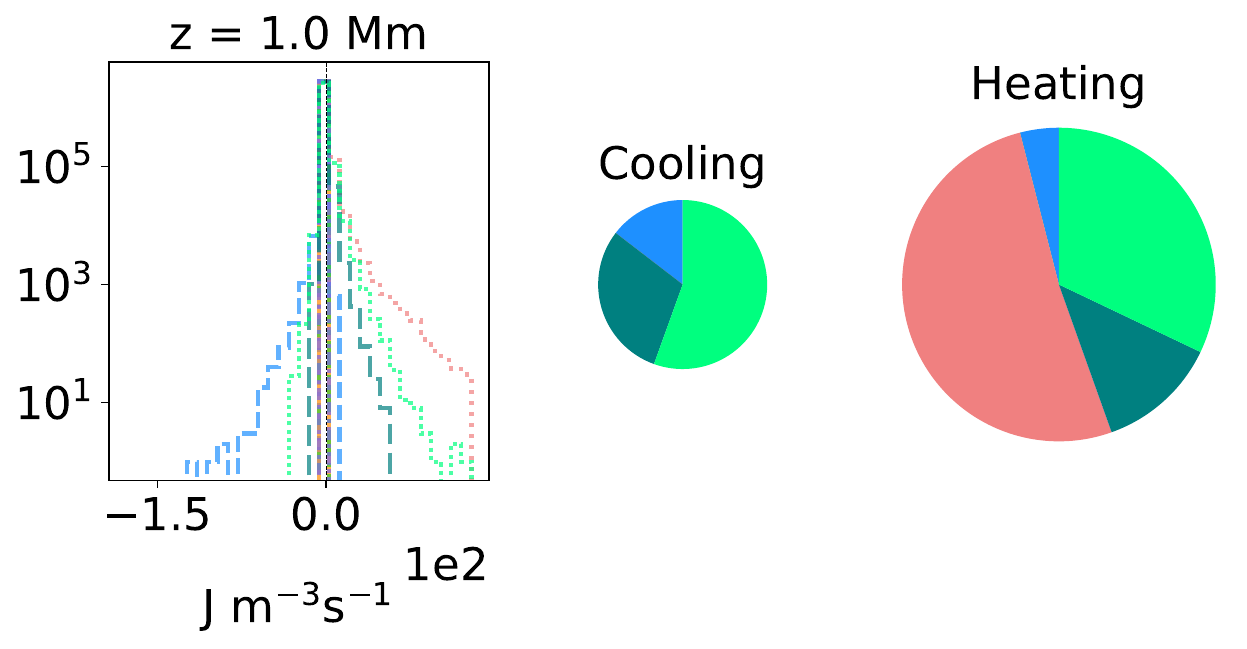}
    \includegraphics[width=0.35\linewidth]{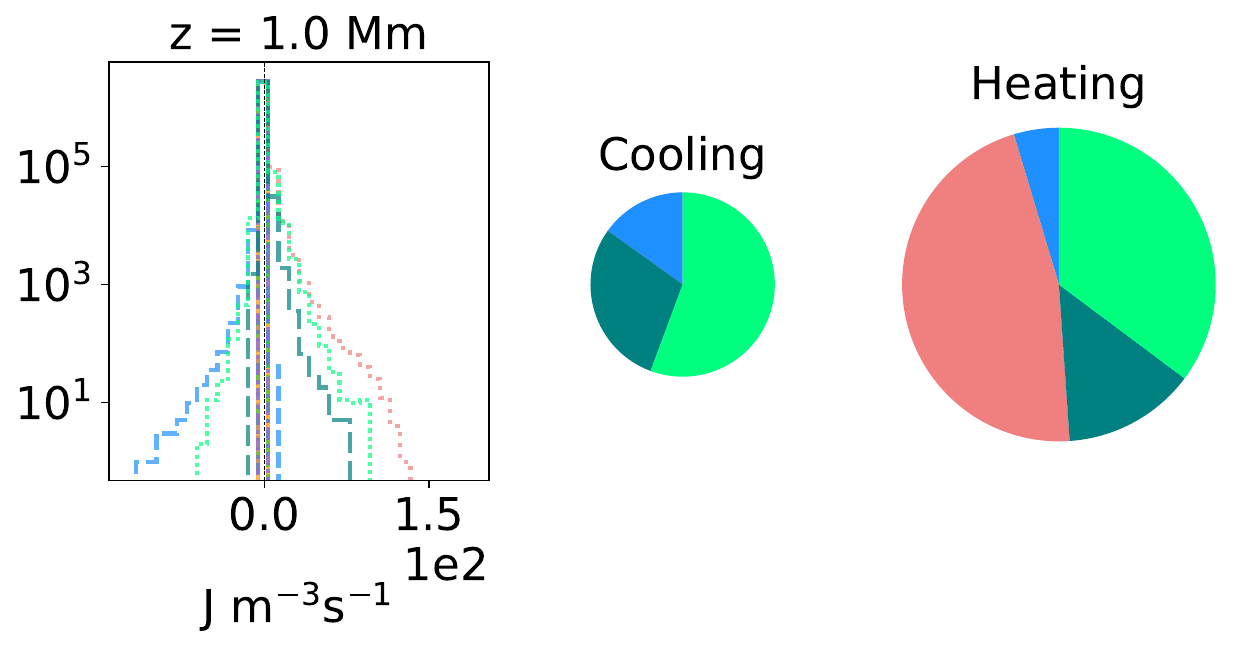}
    \includegraphics[width=0.35\linewidth]{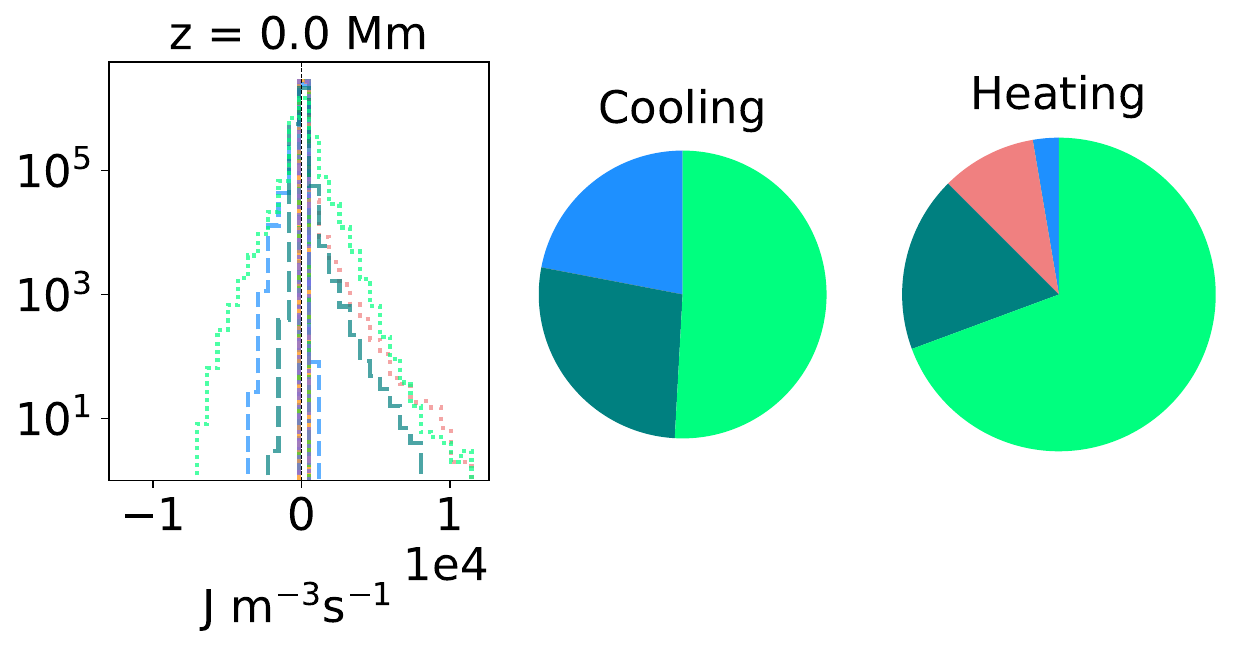}
    \includegraphics[width=0.35\linewidth]{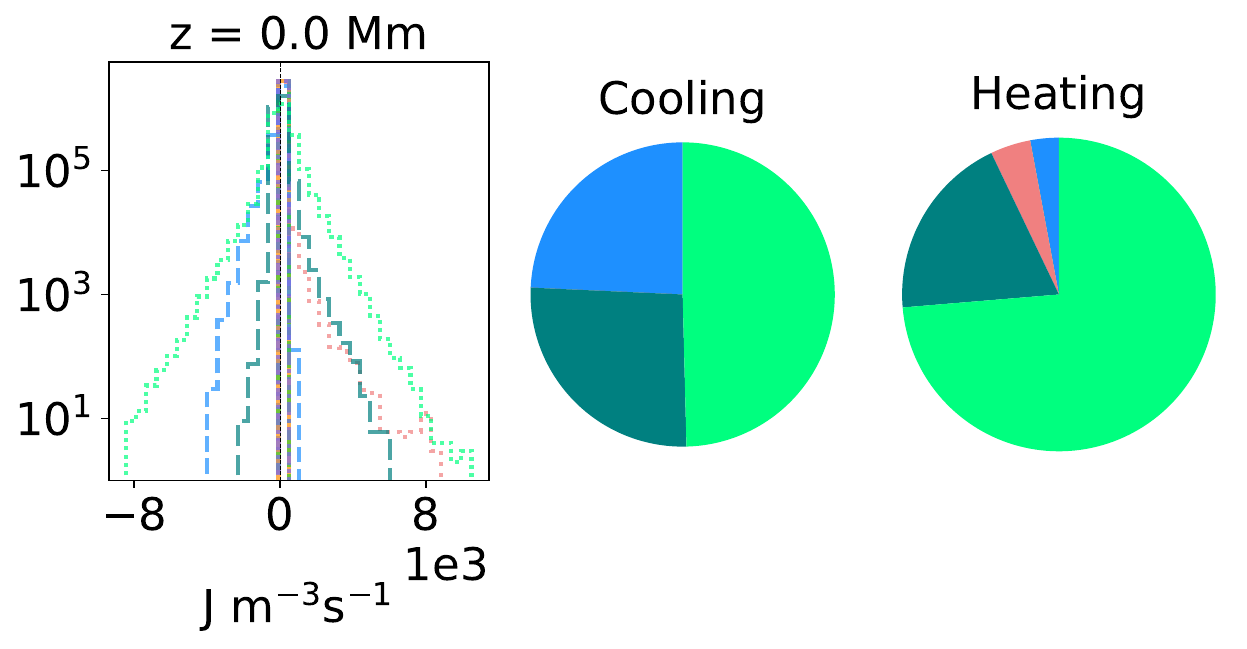}
    \includegraphics[width=0.35\linewidth]{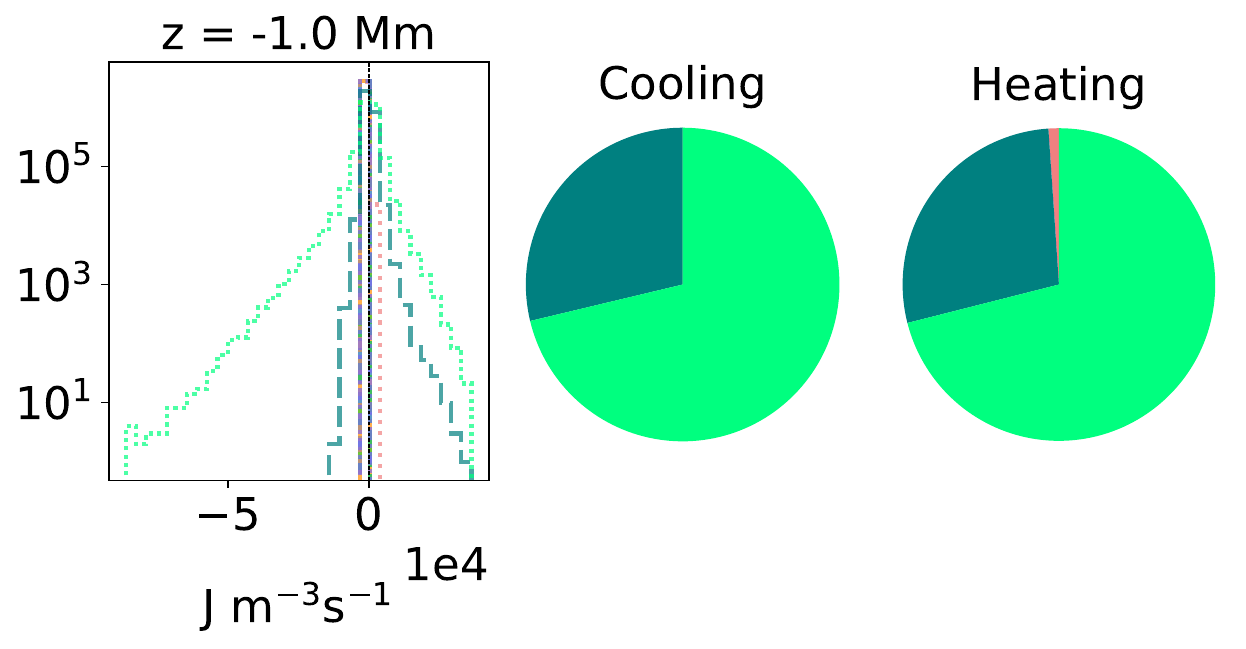}
    \includegraphics[width=0.35\linewidth]{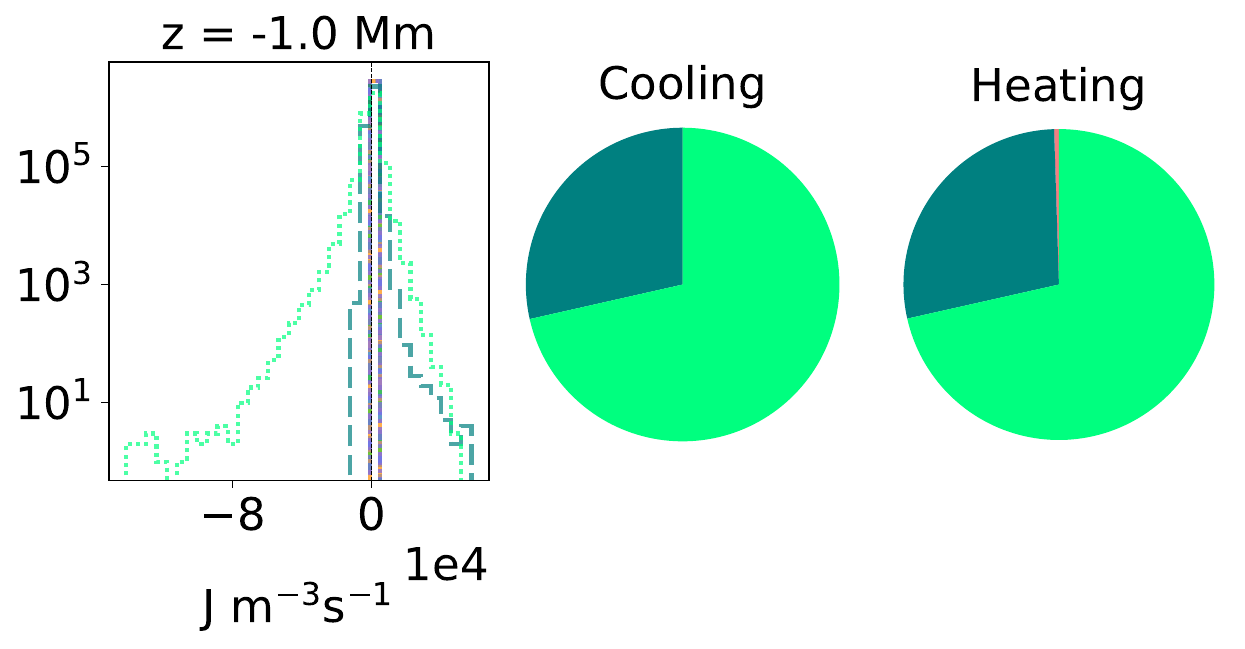}
    \includegraphics[width=0.6\linewidth]{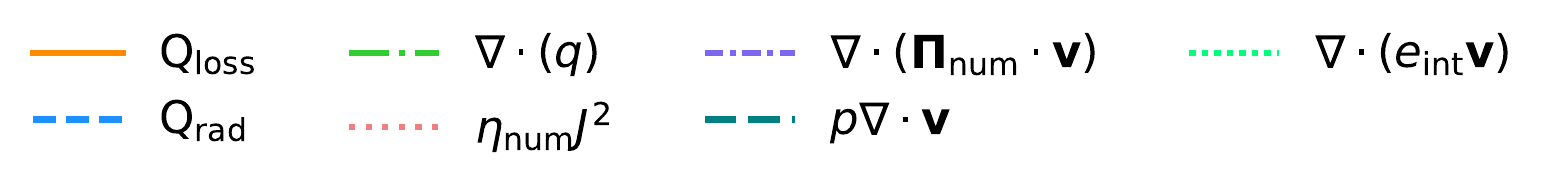}
    \caption{Graphical analysis of the heating and cooling terms from Figure \ref{fig:eint_eq} at selected heights. Histograms and pie charts represent the contributions of the heating and cooling terms in the internal energy equation at $z =$ 10, 6, 3, 1, 0, -1 Mm. The left column of plots corresponds to the simulation with a vertical magnetic field, and the right column to the simulation with an arcade-like magnetic field. Each histogram is accompanied by two pie charts illustrating the relative contributions of the individual terms to the total heating and cooling. }
    \label{fig:pie_charts}
\end{figure*}
\begin{figure}
    \centering
    \includegraphics[width=0.9\linewidth]{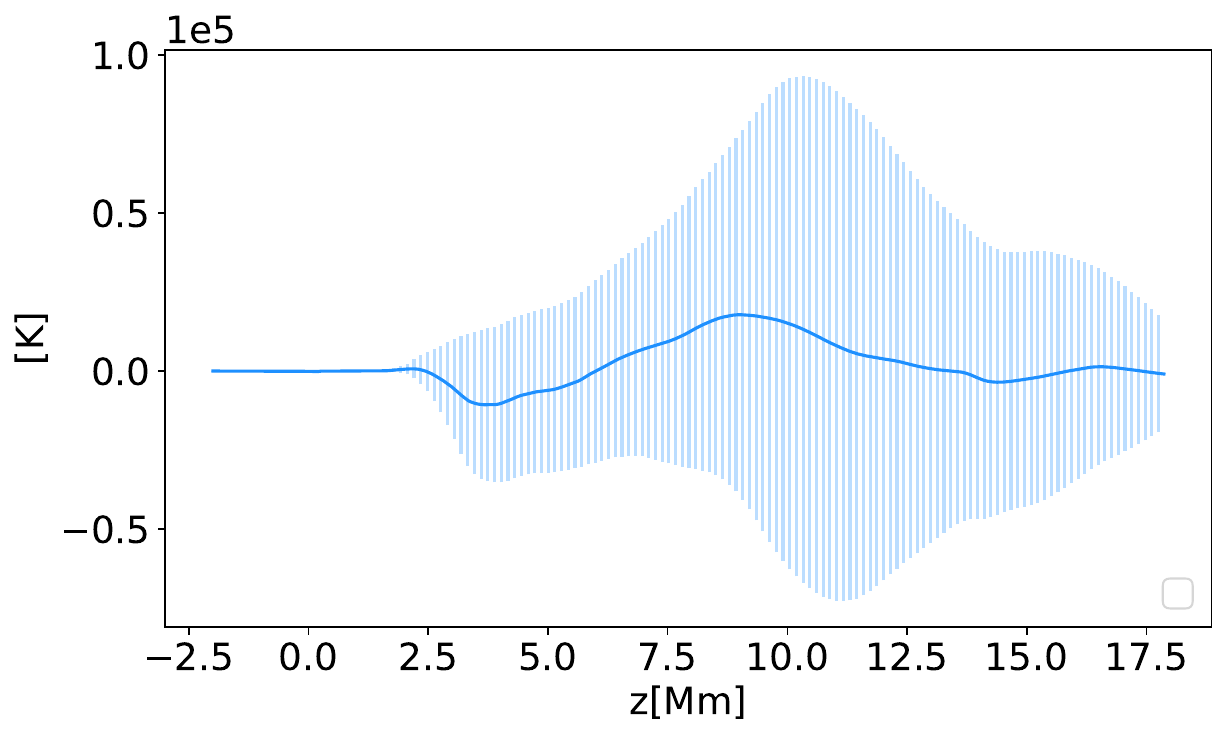}
    \caption{Difference in horizontally and temporally averaged temperature between simulations using Braginskii and Spitzer heat flux models during the final 20 minutes of the simulation. Vertical error bars indicate the propagated standard deviation.}
    \label{fig:average_te_Brag_vs_Sp_delta}
\end{figure}
Figure \ref{fig:eint_eq} shows the horizontally and temporally averaged contributions to the internal energy equation, computed over the final 20 minutes of each simulation. For clarity, values smaller than $1\times10^{-10}$ are omitted from the plots. Overall, the qualitative shape of the total heating and cooling profiles is similar in both simulations. In the case of the vertical magnetic field, the total cooling and heating contributions are nearly identical below a height of $z = 3$ Mm, but in the corona, the total cooling exceeds the heating. This indicates that the boundary condition maintaining the temperature at 1 MK dominates the upper atmosphere; without it, the corona would cool rapidly, as confirmed by independent test runs. In contrast, for the arcade-like magnetic field simulation, heating and cooling contributions are closely balanced at all heights, indicating a more thermally self-regulated state. However, in this case, without a fixed temperature at the top boundary, the corona would also cool, but at a slower rate than in the vertical-field case.

According to Figure \ref{fig:eint_eq}, the dominant heating mechanism throughout the atmosphere is Joule heating resulting from numerical resistivity. Below $z = 3$ Mm, the primary cooling term is the optically thick radiative loss term \qrad. In the corona, the main cooling contributions arise from optically thin radiative losses \qloss, and thermal conduction ($\nabla \cdot \qtc$). The pressure work ($p\nabla\cdot{\bf v}$) term exhibits both heating and cooling behavior in the arcade-like field simulation, depending on height. The contribution from numerical viscous dissipation is present in both heating and cooling terms, but its magnitude is approximately five orders smaller than the dominant contributions.

While averaged contributions provide a general view, they do not always capture the full picture of which terms dominate the thermal structure at specific locations. To address this, Figure \ref{fig:pie_charts} presents a more detailed analysis of the individual contributions to the internal energy equation at selected heights: $z = $ 10, 6, 3, 1, 0, and -1 Mm. Each row of plots corresponds to one of these heights and shows histograms of the different terms in Equation \eqref{eq:evol_eint}, computed over the final 20 minutes of each simulation. The left panels correspond to the simulation with a vertical magnetic field, and the right panels to the arcade-like magnetic field.  Next to each histogram are two pie charts that illustrate the relative contributions of the individual terms to the total heating and total cooling, respectively. The size of each pie chart is proportional to the total heating or cooling at that height. This figure offers a more comprehensive view of the internal energy exchange processes at different layers of the atmosphere and highlights the relative importance of each physical mechanism.

The detailed analysis in Figure \ref{fig:pie_charts} reveals key differences in the heating and cooling balance across atmospheric layers. In fact, very few terms are visible in each of these plots.  A comparison between the two simulations reveals that the flux contributions are generally similar across all heights. However, in the coronal region -at heights $z = $6 and 10 Mm, the balance between heating and cooling is more evenly distributed in the arcade configuration. In contrast, the vertical field case exhibits slightly more cooling than heating at those altitudes. 

Radiative cooling becomes particularly significant in the mid-corona around $z = 6$ Mm, but its influence diminishes at higher altitudes. Thermal conduction, on the other hand, plays a dominant role in both heating and cooling throughout the coronal heights, contributing substantially to the energy exchange processes. 

At $z = 1$ Mm, there is a net heating effect, where the heating rate exceeds the cooling rate by a factor of $\sim$1.8, where the most dominant heating mechanism is the joule heating due to numerical resistivity.  Only at these heights is the contribution of the numerical resistivity important in magnitude to the other terms, rather than uniformly throughout the atmosphere as Figure \ref{fig:eint_eq} might suggest, where its dominance emerged from the mutual cancellation of other terms in the average.

In the lower atmosphere, the radiative term for optically thin plasma (\qrad) is most significant at photospheric and chromospheric heights.  In almost all heights, the internal energy flux and the pressure work term are key contributors to both heating and cooling and are especially important for near-surface convection layers.

\subsubsection{Effects of the perpendicular thermal conductivity}

The thermal conductivity model used in our simulations is based on the formulation by Braginskii, as discussed in Section \ref{subsec:tc}. However, many similar studies adopt the model developed by Spitzer, which neglects the perpendicular component of thermal conduction. Although this simplification is often justified, we hypothesize that the perpendicular component may still play a non-negligible role in energy transport under certain conditions. To assess the significance of this effect, we conducted Simulation 4, which uses the Spitzer thermal conduction model. This simulation employs the same setup as the arcade-like magnetic field configuration of Simulation 3. Animations illustrating the time evolution of density and temperature for Simulation 4 are provided in Table \ref{tab:description}. Overall, the simulation behaves similarly to Simulation 3, though some differences are worth noting.

Figure \ref{fig:average_te_Brag_vs_Sp_delta} shows the difference in horizontally and time-averaged temperature between simulations 3 and 4 over the final 20 minutes of the runs. The vertical bars represent the propagated standard deviation, providing an estimate of the uncertainty. Although the differences may not be statistically significant, they suggest that distinct physical behavior is occurring between the two simulations.
\begin{figure}
    \centering
    \includegraphics[width=0.49\textwidth]{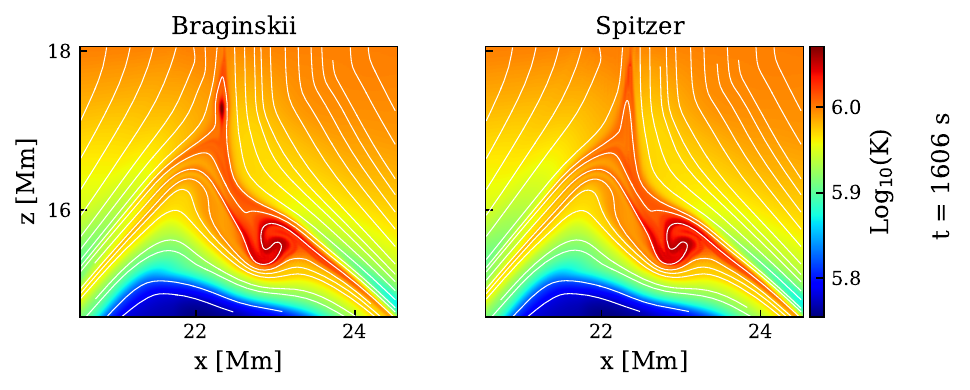} \\
     \includegraphics[width=0.49\textwidth]{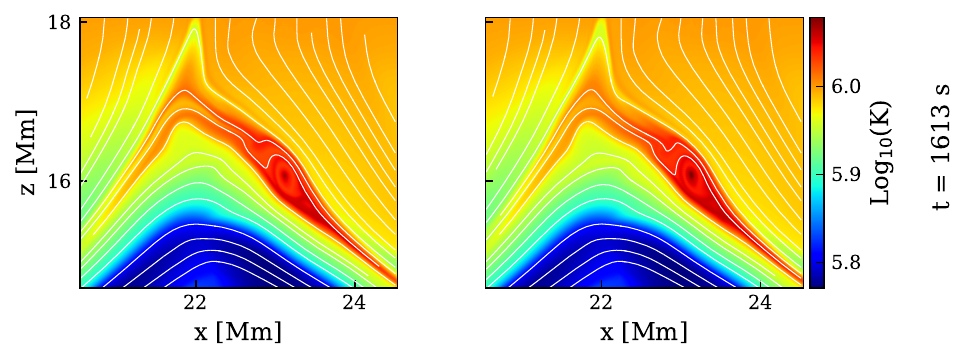} \\
     \includegraphics[width=0.49\textwidth]{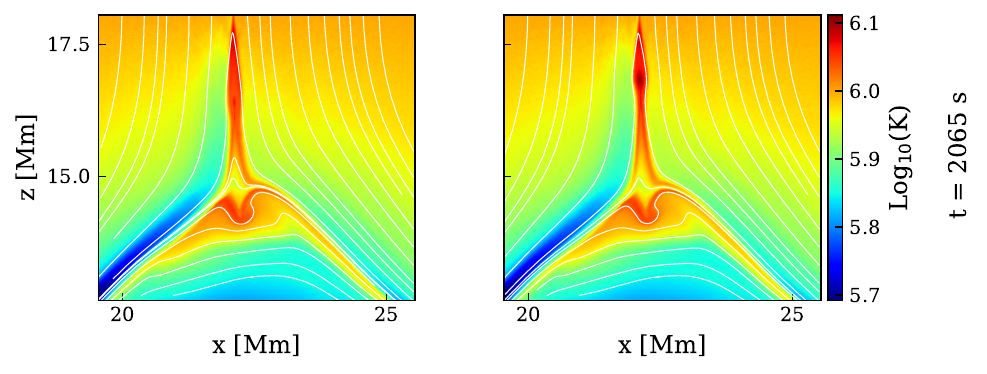}\\
     \includegraphics[width=0.49\textwidth]{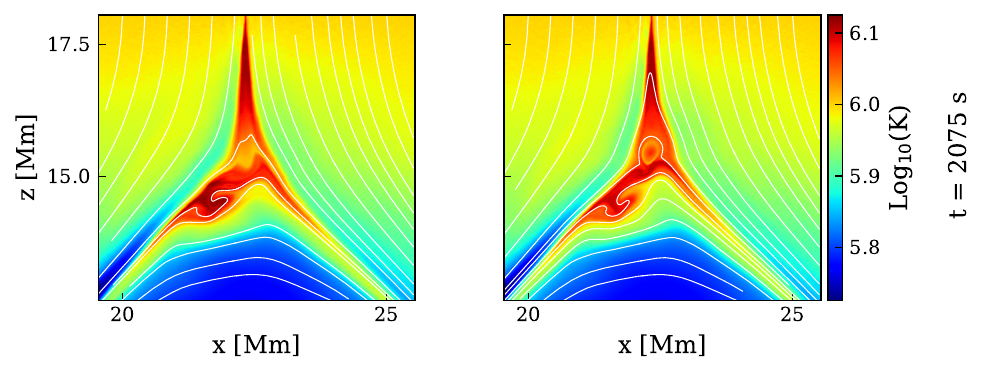}\\
      \includegraphics[width=0.49\textwidth]{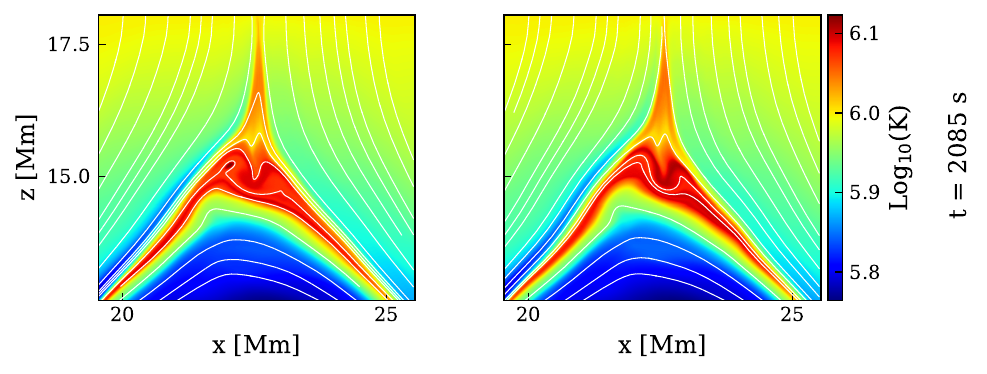}
    \caption{Comparison of the temperature colormap at selected snapshots for simulations with different heat flux models. The left column corresponds to the Braginskii model and the right to the Spitzer model. Each row represents a different time: from top to bottom —t=1606 s, t=1613 s, t=2065 s, t=2075 s, and t=2085 s.}
    \label{fig:colormaps_Brag_vs_Sp}
\end{figure}

To further check on the origin of these differences, Figure \ref{fig:colormaps_Brag_vs_Sp} compares temperature colormaps between the two simulations in selected zoomed-in regions at different times. The first column of plots corresponds to the simulation with the Braginskii model and the right column to the Spitzer model. Each row represents a snapshot at a different simulation time: from top to bottom-$t=$1606, 1613, 2065, 2065, 2075, and 2085 s.

These comparisons reveal localized differences in temperature structure between the two models during magnetic reconnection events near the top boundary, where null points are formed.  Although individually small, the cumulative effect of many such events likely explains the differences seen in the time-averaged temperature profile in Figure \ref{fig:average_te_Brag_vs_Sp_delta}. 

From an analytical point of view, this behavior is expected. Eq. \ref{eq:kappa_perp_e} shows how the perpendicular component of thermal conduction becomes comparable to the parallel component when the magnetic field strength approaches zero, as occurs at null points formed during reconnection. Therefore, while the impact of the perpendicular conductivity is limited in most regions, it becomes relevant in localized areas of magnetic reconnection.

\section{Discussion and conclusions} \label{sec:conclusions}

We have introduced the new MAGEC code, providing a detailed description of its numerical methods, as well as its physical modules for thermal conduction and coronal radiative losses. We have also presented the methodology used to estimate the effective numerical resistivity and viscosity. To evaluate the performance and accuracy of the code, we carried out a series of simulations of the solar atmosphere. These simulations serve to demonstrate the robustness and capabilities of MAGEC as a reliable and versatile numerical tool for solar physics research.

In these simulations, we investigated how the inclination of the magnetic field influences the thermal structure of the solar atmosphere. Our results show that a vertical magnetic field configuration sustains a hotter corona at intermediate heights compared to an arcade-like configuration. The stability of our models achieved by the inclusion of a hot plate at the boundary top reinforces the well-established conclusion that an additional heating mechanism is required to maintain coronal temperatures.  These findings are consistent with results obtained from other widely used simulation codes.

Through a detailed analysis of the energy equation—including both explicit and implicit  contributions—we found that, although numerical resistivity appears dominant in time-averaged heating profiles, its effects are concentrated primarily in the photosphere and chromosphere. In contrast, within the corona, the dominant energy transport mechanisms are thermal conduction and pressure work. The interplay of various heating and cooling processes across different atmospheric layers leads to a complex thermal structure, with alternating regions of net heating and net cooling.

We also examined the role of the perpendicular component of thermal conduction, which is often neglected in coronal models. By comparing simulations using the Braginskii and Spitzer conductivity formulations, we found that while the overall temperature profiles are broadly similar, the perpendicular component can influence local plasma dynamics, especially in the vicinity of magnetic reconnection events. When such events occur frequently, the cumulative effect of these localized differences can become apparent in the average temperature profile.

Although numerous MHD codes have been developed for solar physics, and several of them employ HRSC methods, we are some of the first to use these techniques to model both the near-surface solar interior and the atmosphere in the same computational box. Our approach uniquely incorporates key coronal physical processes, such as optically thin radiative losses and anisotropic thermal conduction, enabling more realistic and self-consistent modeling of the coupled solar interior–corona system.  In future work, we plan to extend its capabilities to fully three-dimensional simulations and incorporate additional physical effects, including ambipolar diffusion, the Hall effect, and the Biermann battery.

\section{Acknowledgments}

This research was funded by the Agencia Estatal de Investigación del Ministerio de Ciencia, Innovación y Universidades (MCIU/AEI) and the European Regional Development Fund (ERDF) under grant "MHD MODeling and beyond: in preparation for the European Solar Telescope"   with reference PID2021-127487NB-I00 and by the MCIU/AEI and the European Social Fund (ESF) under grant "Ayudas Ramón y Cajal Convocatoria 2020" with reference RYC2020-030307-I. It also contributes to the deliverables outlined in the FP7 European Research Council grant agreement ERC-2017-CoG771310-PI2FA for the project “Partial Ionization: Two-fluid Approach”. TF acknowledges support from the Project CNS2023-145233, funded by MICIU/AEI/10.13039/501100011033, and the European Union “NextGenerationEU/RTRP”. 
AN gratefully acknoweledges financial support from grant PID2024-156538NB-I00, funded by MCIN/AEI/ 10.13039/501100011033 and by “ERDF A way of making Europe”. The authors also acknowledge the technical support and expertise provided by the Spanish Supercomputing Network (Red Española de Supercomputación). Computational resources were provided by the LaPalma Supercomputer at the Instituto de Astrofísica de Canarias and the MareNostrum Supercomputer in Barcelona, Spain.

\bibliographystyle{aa}



\begin{appendix}
\section{The numerical methods \label{appendix}}

This section outlines the numerical methods employed for the 1D case, with straightforward extensions to the 3D case achieved through cyclic permutations of the variables. The expressions are provided in SI units for clarity and consistency with our code implementation.

\subsection{Reconstruction of the variables}

In finite volume methods, the fluxes of conserved quantities (e.g, mass, momentum, energy) are computed at cell interfaces. However, the state vector $U_{i}$ is defined over the entire cells, not the specific values at interfaces. To address this, High Resolution Capturing (HRSC) methods reconstruct the state vector at the left $U^L_{i+1/2}$ and right $U^R_{i+1/2}$ sides of each interface. This reconstruction converts the state vector into a collection of local Riemann problems.

For instance, a piecewise linear reconstruction yields
\begin{eqnarray}
U^L_{i+1/2} = U_i + \frac{1}{2}\sigma_i  ,  \ \  U^R_{i+1/2} = U_{i+1} - \frac{1}{2}\sigma_{i+1} ,    \label{eq:left_right_states}
\end{eqnarray}
where $\sigma_i$ represents the slope of the solution in cell $i$. A commonly used slope limiter is the minmod limiter, 
\begin{equation}
\sigma_i = \text{minmod}(U_i - U_{i-1}, U_{i+1} - U_i) ,    
\end{equation}
 with the minmod function defined as:
\begin{equation}
\text{minmod}(a,b) = \left\lbrace \begin{array}{l}
    a, \ \  \text{if} \ \  |a|<|b| \ \text{and} \ ab > 0 , \\
    b, \ \ \text{if} \ \ |b|<|a| \ \text{and} \ ab > 0 ,  \\
    0, \ \ \text{if} \ \ ab \leq 0 .
\end{array}  \right.
\end{equation}
In addition to the piecewise linear method (PLM), more sophisticated reconstruction techniques can be employed, such as the Piecewise Parabolic methods (PPM), or Weighted Essentially Non-Oscillatory (WENO) methods. In the code we have implemented the options MINMOD \citep{minmod}, MC \citep{VANLEER1977263}, Van Leer \citep{VANLEER1974361}, Ospre \citep{ospre}, and WENO5 \citep{TitarevToro2004}.

\subsection{HLL Riemann solver}

Once a Riemann problem is defined on both sides of every intercell, the HRSC methods use a Riemann solver to calculate fluxes at cell interfaces. To simplify the notation, the left and right states calculated in eq \ref{eq:left_right_states} are denoted as ${\bf U}_\text{L}$ and ${\bf U}_\text{R}$, respectively.  In the code we have implemented the HLL \citep{Harten_etal_1983}, HLLC \citep{LI2005344}, and HLLD \citep{MIYOSHI2005315} Riemann Solvers, each offering different levels of accuracy and stability. Below, we present the expressions for these solvers. 

The structure of the solution of the Riemann problem (Riemann fan) is bounded by the slowest and fastest signal speeds of the system \eqref{eq:eigenvalues}, 
\begin{eqnarray}
 S_\text{L} = \mathrm{min} \left[ \lambda_i^R, \lambda_i^L \right] \, , \,  S_\text{R} = \mathrm{max} \left[ \lambda_i^R, \lambda_i^L \right] \, . 
\end{eqnarray}
\begin{figure}
    \centering
    \includegraphics[width=0.7\linewidth]{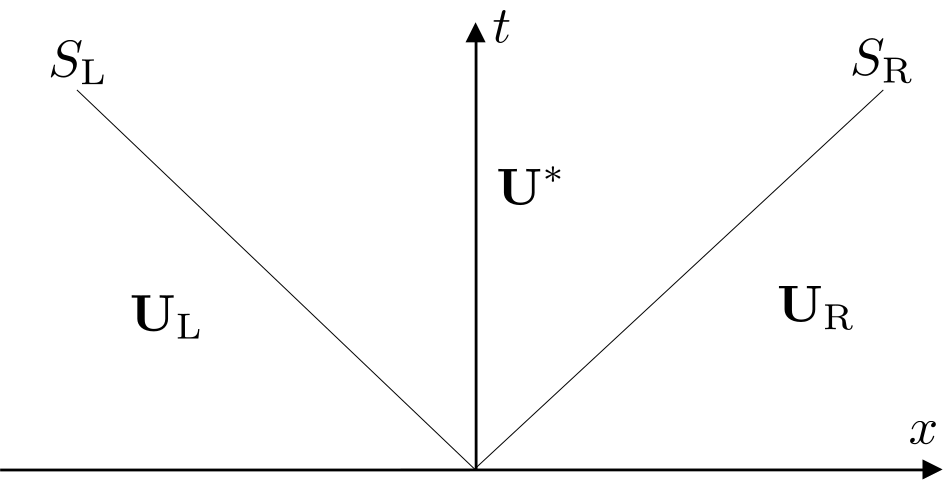}
    \caption{Schematic diagram of the Riemann fan for the HLL Riemann solver. The exact wave fan is approximated by two waves, $S_\text{L}$ and $S_\text{R}$, with a uniform state (${\bf U^*}$) in between. }
    \label{fig:hlle_riemann_fan}
\end{figure}
The HLL (Harten-Lax-van Leer) Riemann solver \citep{Harten_etal_1983} is designed by assuming an average intermediate state between the fastest and slowest waves, as shown in figure \ref{fig:hlle_riemann_fan}. This state, denoted ${\bf U^*}$ is given by 
\begin{eqnarray}
{\bf U}^* = \frac{S_\text{R} {\bf U}_\text{R} - S_\text{L} {\bf U}_\text{L} + {\bf F}_\text{L} - {\bf F}_\text{R}}{S_\text{R} - S_\text{L}} ,
\end{eqnarray}
where ${\bf F}_\text{L}$ and ${\bf F}_\text{R}$ are the fluxes evaluated at the left and right states, ${\bf F}_\text{L}={\bf F}({\bf U}_\text{L})$ and ${\bf F}_\text{R}={\bf F}({\bf U}_\text{R})$. The estimated flux ${\bf F}_{\text{HLL}}$, is determined as   
\begin{eqnarray}
    {\bf F}_{\text{HLL}} = \left \lbrace \begin{array}{ccl}
    {\bf F}_\text{L} & \text{if} &  S_\text{L} >0,  \\
    {\bf F}^* & \text{if} &  S_\text{L} \leq 0 \leq S_\text{R} , \\
    {\bf F}_\text{R}  & \text{if} &  S_\text{R}<0. \\
    \end{array}
    \right. ,
\end{eqnarray}
where ${\bf F}^*$ is the flux corresponding to the intermediate state ${\bf U}^*$, is given by 
\begin{equation}
{\bf F}^* = \frac{  S_\text{R}  {\bf F}_\text{L}  - S_\text{L}  {\bf F}_\text{R}  + S_\text{R}  S_\text{L}  \left({\bf U}_\text{R}  - {\bf U}_\text{L} \right)}{S_\text{R} -S_\text{L} }  \, .  
\end{equation}
This method is robust, simple, and computationally efficient since it does not require resolving the full wave structure of the Riemann problem, and for the same reason, can be over diffusive. 

\subsection{HLLC Riemann solver}

\begin{figure}
    \centering
    \includegraphics[width=0.7\linewidth]{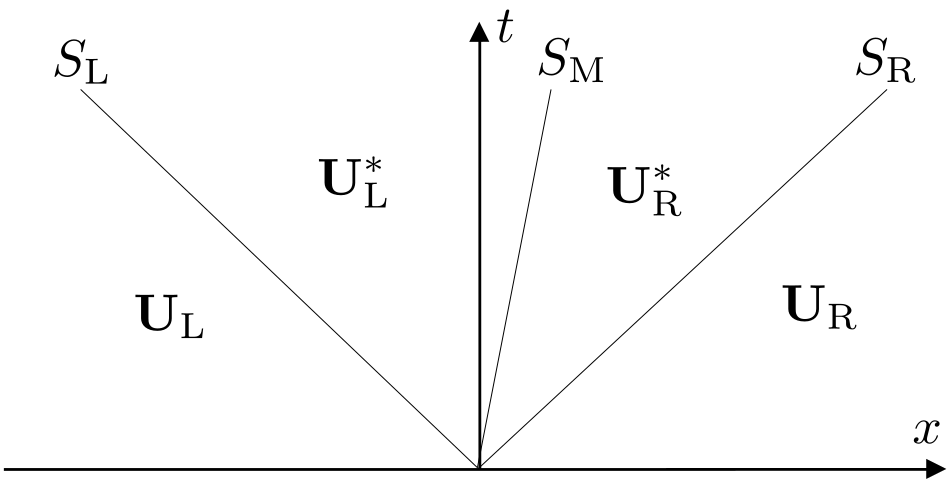}
    \caption{Schematic diagram of the Riemann fan for the HLLC Riemann solver. The exact wave fan is approximated by three waves, $S_\text{L}$, $S^*$, and $S_\text{R}$, with two uniform states (${\bf U_\text{L}^*}$ and ${\bf U_\text{R}^*}$) separated by a contact discontinuity.}
    \label{fig:hllc_riemann_fan}
\end{figure}

The HLLC (Harten-Lax-van Leer-Contact) method, proposed for hydrodynamics by \cite{Toro_hllc} and has been adapted for MHD by different authors. It improves the HLL solver by resolving an additional middle wave associated with the contact discontinuity. This creates left ${\bf U}^*_\text{L}$ and right ${\bf U}^*_\text{R}$ intermediate states, as illustrated in Fig. \ref{fig:hllc_riemann_fan}. We use the MHD version derived by \cite{LI2005344}.

 Assuming that the pressure and velocities do not change across the contact discontinuity, we have 
\begin{eqnarray}
& & p^{*}_{\text{L}} = p^{*}_{\text{R}} = p^*  , \\ 
& & v^{*}_{x\text{L}} = v^{*}_{x\text{R}} = v_x^{*} = S_{\text{M}}  .  
\end{eqnarray} 
Here $p^{*}_{\text{L}}$ and $v^{*}_{x\text{L}}$ denote the presssure and x-velocity compontent associated to the state ${\bf U^*_\text{L}}$, while $p^{*}_{\text{R}}$ and $v^{*}_{x\text{R}}$ correspond to the same quantities for the state ${\bf U^*_\text{R}}$. $S_\text{M}$ represents the velocity of the contact wave and is given by 
\begin{eqnarray}
& & S_\text{M} = \frac{\rho_\text{R} v_{x\text{R}} (S_\text{R} - v_{x\text{R}}) - \rho_\text{L} v_{x\text{L}} (S_\text{L} - v_{x\text{L}})}
{\rho_\text{R} (S_\text{R} - v_{x\text{R}}) - \rho_\text{R} (S_\text{R} - v_{x\text{L}}) } \\
& & \hspace{0.9cm} + \frac{p_\text{L} - p_\text{R} }
{\rho_\text{R} (S_\text{R} - v_{x\text{R}}) - \rho_\text{R} (S_\text{R} - v_{x\text{L}})} \, . 
\end{eqnarray}
The flux ${\bf F}_{\text{HLLC}}$ is given by
\begin{eqnarray}
    {\bf F}_{\text{HLLC}} = \left \lbrace \begin{array}{ccl}
    {\bf F}_\text{L}  & \text{if} &  S_\text{L} >0,  \\
    {\bf F}^*_\text{L}  & \text{if} &  S_\text{L}  \leq 0 \leq S_\text{M} , \\
    {\bf F}^*_\text{R}  & \text{if} &  S_\text{M}  \leq 0 \leq S_\text{R} , \\
    {\bf F}_\text{R}  & \text{if} &  S_\text{R} <0, \\
    \end{array}
    \right. 
\end{eqnarray}
where
\begin{eqnarray}
& & {\bf F}^*_\text{L} = {\bf F}_\text{L} + S_\text{L} \left(  {\bf U}^*_\text{L} - {\bf U}_\text{L} \right) \, , \\
& & {\bf F}^*_\text{R} = {\bf F}_\text{R} + S_\text{R} \left(  {\bf U}^*_\text{R} - {\bf U}_\text{R} \right) \, , 
\end{eqnarray}
and the intermediate state variables are given by
\begin{eqnarray}
& & p^*  =  \rho_\text{L} (S_\text{L} - v_x)(S_\text{M} - v_x) + p_\text{L} - \frac{B_x^2}{\mu_0} + \frac{{B_x^*}^2}{\mu_0}  \, , \\
& & {B_i^*}_\text{L} = {B_i^*}_\text{R} = B_i^* = B_i^\text{HLL} \, , \\
& & \rho^*_\text{L} = \rho_\text{L}\frac{S_\text{L}-v_x}{S_\text{L} -S_\text{M}} \, , \\
& & \left( \rho v_x \right)^*_\text{L} = \rho^*_\text{L} S_\text{M} \, , \\
& & \left( \rho v_y \right)^*_\text{L} = (\rho v_y)_\text{L} \frac{S_\text{L}-v_x}{S_\text{L}- S_\text{M}} - \frac{ B^*_xB^*_y - B_xB_y }{(S_\text{L} - S_\text{M})\mu_0} \, , \\
& & \left( \rho v_z \right)^*_\text{L} = (\rho v_z)_\text{L} \frac{S_\text{L}-v_x}{S_\text{L}- S_\text{M}} - \frac{ B^*_xB^*_z - B_xB_z }{(S_\text{L} - S_\text{M})\mu_0} \, , \\
& & e^*_\text{L} = \frac{ e(S_\text{L}-v_x) + p^*S_\text{M}-p v_x  }{S_\text{L} - S_\text{M}} \nonumber \,  \\ 
& &  \hspace{0.8cm}   - \frac{\left[ B^*_x( {\bf B} \cdot {\bf v} )^* - B_x( {\bf B}\cdot {\bf v} ) \right] }{ (S_\text{L} - S_\text{M} )\mu_0  }   \, .
\end{eqnarray}
The expressions for the right state are analogous to those for the left state, with L replaced by R.

This method offers enhanced accuracy by explicitly resolving the contact discontinuity, which reduces the over-diffusive nature of the basic HLL solver. It achieves a balance between computational efficiency and precision, making it suitable for many hydrodynamic and MHD applications.

\subsection{HLLD Riemann solver}

\begin{figure}
    \centering
    \includegraphics[width=0.7\linewidth]{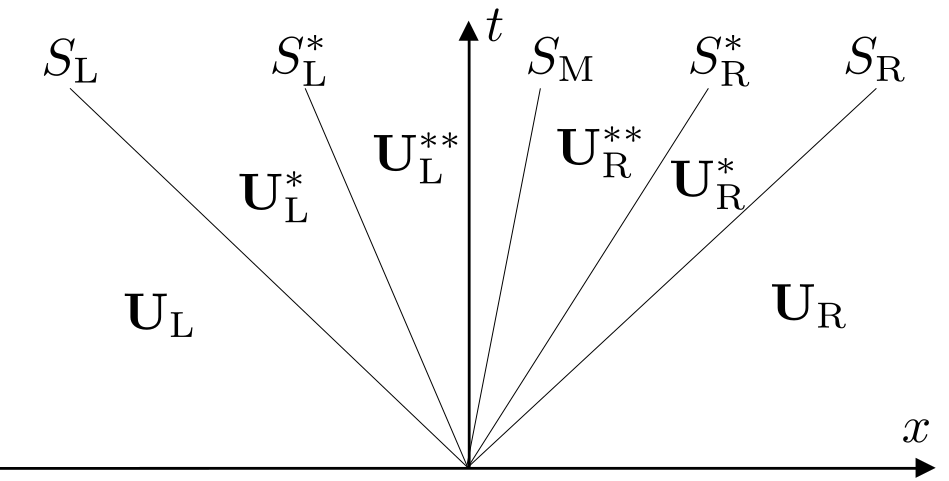}
    \caption{Schematic diagram of the Riemann fan for the HLLD Riemann solver. The exact wave fan is approximated by five waves—$S_\text{L}$, $S_\text{L}^*$, $S_\text{M}$, $S_\text{R}^*$ and $S_\text{R}$— with four constant states ({\bf $U_\text{L}^*$}, {\bf $U_\text{L}^{**}$}, {\bf $U_\text{R}^{**}$} and {\bf $U_\text{R}^{*}$}) separated by Alfvén and contact discontinuities. }
    \label{fig:hlld_riemann_fan}
\end{figure}

The HLLD \citep{2005JCoPh.208..315M} Riemann solver extends the capabilities of the HLLC solver by introducing a more detailed structure of the Riemann fan. It divides the solution into four intermediate states  ${\bf U}^*_\text{L}$, ${\bf U}^*_\text{R}$, ${\bf U}^{**}_\text{L}$ and ${\bf U}^{**}_\text{R}$, as shown in Fig. \ref{fig:hlld_riemann_fan}
The flux ${\bf F}_{\text{HLLD}}$ is given by
\begin{eqnarray}
    {\bf F}_{\text{HLLD}} = \left \lbrace \begin{array}{ccl}
    {\bf F}_\text{L}  & \text{if} &  S_\text{L} >0,  \\
    {\bf F}^*_\text{L}  & \text{if} &  S_\text{L}  \leq 0 \leq S^*_\text{L} , \\
    {\bf F}^{**}_\text{L}  & \text{if} &  S^*_\text{L}  \leq 0 \leq S_\text{M} , \\
    {\bf F}^{**}_\text{R}  & \text{if} &  S_\text{M}  \leq 0 \leq S^*_\text{R} , \\   
    {\bf F}^*_\text{R}  & \text{if} &  S^*_\text{R}  \leq 0 \leq S_\text{R} , \\
    {\bf F}_\text{R}  & \text{if} &  S_\text{R} <0, \\
    \end{array}
    \right. 
\end{eqnarray}
where the intermediate fluxes ${\bf F}^{**}_\text{L}$ and ${\bf F}^{**}_\text{R}$ are calculated as
\begin{eqnarray}
& & {\bf F}^{**}_\text{L} = {\bf F}^*_\text{L} + S^*_\text{L} \left(  {\bf U}^{**}_\text{L} - {\bf U}^*_\text{L} \right) \, , \\
& & {\bf F}^{**}_\text{R} = {\bf F}^*_\text{R} + S^*_\text{R} \left(  {\bf U}^{**}_\text{R} - {\bf U}^*_\text{R} \right) \, .  
\end{eqnarray}
The intermediate wave velocities are defined as  
\begin{eqnarray}
& & S_\text{M} = \frac{\rho_\text{R} v_{x\text{R}} \left(S_\text{R} - v_{x\text{R}}\right) - \rho_\text{L} v_{x\text{L}} (S_\text{L} - v_{x\text{L}})}
{\rho_\text{R} (S_\text{R} - v_{x\text{R}}) - \rho_\text{R} (S_\text{R} - v_{x\text{L}}) } \\
& & \hspace{0.8cm} + \frac{p_{T\text{L}} - p_{T\text{R}} }
{\rho_\text{R} (S_\text{R} - v_{x\text{R}}) - \rho_\text{R} (S_\text{R} - v_{x\text{L}})},  \\
& & S^*_\text{L} = S_\text{M} - \frac{|B_x|}{\sqrt{\mu_0\rho^*_\text{L}}} \, , \ S^*_\text{R} = S_\text{M} + \frac{|B_x|}{\sqrt{\mu_0\rho^*_\text{R}}} \, , 
\end{eqnarray}
where $p_{T\text{L}}$ and $p_{T\text{R}}$ are the total pressures on the left and right states. Additionally, the normal velocity and total pressure are assumed to be constant over the Riemann fan:
\begin{eqnarray}
  & & v^*_{x\text{L}}  = v^{**}_{x\text{L}} = v^{**}_{x\text{R}} = v^{*}_{x\text{R}}  = S_\text{M}  \, , \\ 
  & & p^*_{T\text{L}}  = p^{**}_{T\text{L}} = p^{**}_{T\text{R}} = p^{*}_{T\text{R}}  = p_T^*  \, . 
\end{eqnarray}
The variables of the state vector ${\bf U}^*_\text{L}$ and ${\bf U}^*_\text{R}$ are computed as follows
\begin{align}
 \rho^*_\text{L} &= \rho_\text{L}\frac{S_\text{L}-v_{x\text{L}}}{ S_\text{L} - S_\text{M} }   \\
 v^*_{y\text{L}} &= v_{y\text{L}} - \frac{B_x B_{y\text{L}} (S_\text{M} - v_{x\text{L}} ) } { \left[ \rho_\text{L}  (S_\text{L} - v_{x\text{L}} )(S_\text{L} - S_\text{L}) - B_x^2 \right] \mu_0 }, \\
 v^*_{z\text{L}} &= v_{z\text{L}} - \frac{ B_x B_{z\text{L}} (S_\text{M} - v_{x\text{L}} )}{ \left[ \rho_\text{L} (S_\text{L} - v_{x\text{L}} )(S_\text{L} - S_\text{M} ) - B_x^2 \right]\mu_0 }, \\
 B^*_{y\text{L}} &= B_{y\text{L}} \frac{ \rho_\text{L} (S_\text{L} - v_{x\text{L}} )^2\mu_0 - B_x^2}{\rho_\text{L} (S_\text{L} - u_\text{L} )(S_\text{L} - S_\text{L} )\mu_0 - B_x^2}, \\ 
 B^*_{z\text{L}} &= B_{z\text{L} } \frac{ \rho_\text{L} (S_\text{L} - v_{x\text{L}})^2\mu_0 - B_x^2}{\rho_\text{L} (S_\text{L} - v_{x\text{L}})(S_\text{L} - S_\text{L})\mu_0 - B_x^2} \, , \\
 e^*_\text{L} &= \frac{(S_\text{L} - v_{x\text{L}}) e_\text{L} - p_{T\text{L}} v_{x\text{L}} + p^*_T S_\text{M} }{S_\text{L} - S_\text{M}}  \\ 
&  \hspace{0.8cm} + \frac{B_x (\bf{v}_\text{L} \cdot {\bf B}_\text{L} - {\bf v}^*_{\text{L}} \cdot {\bf B}^*_\text{L} ) }{(S_\text{L} - S_\text{M})\mu_0 } \, , \\
 p^{*}_{T} &= \frac{(S_\text{R} - v_{x\text{R}})\rho_\text{R} p_{T\text{L}} - (S_\text{L} - v_{x\text{L}})\rho_\text{L} p_{T\text{R}} }{(S_\text{R} - v_{x\text{R}})\rho_\text{R} - (S_\text{L} - v_{x\text{L}} )\rho_\text{L} }  \\
& \hspace{0.8cm} + \frac{\rho_\text{L} \rho_\text{R} (S_\text{R} - v_{x\text{R}} )(S_\text{L} - v_{x\text{L}} )(v_{x\text{R}} - v_{x\text{L}} )}
{(S_\text{R} - v_{x\text{R}})\rho_\text{R} - (S_\text{L} - v_{x\text{L}} )\rho_\text{L} }.    
\end{align}
For the states ${\bf U}^{**}_{\text{L}}$ and ${\bf U}^{**}_{\text{R}}$ the following assumptions hold constant across the Riemann fan
\begin{align}
  \rho^{**}_\text{L} &= \rho^*_\text{L} \, , \ p^{**}_{T\text{L}} = p^*_{T\text{L}} \, , \\
  v^{**}_{y\text{L}} &= v^{**}_{y\text{R}} = v^{**}_y   , \   v^{**}_{z\text{L}} = v^{**}_{z\text{R}} = v^{**}_z \, , \\ 
  B^{**}_{y\text{L}} &= B^{**}_{y\text{R}} = B^{**}_y   , \    B^{**}_{z\text{L}} = B^{**}_{z\text{R}} = B^{**}_z  \, ,
\end{align}
where
\begin{align}
 v^{**}_y &= \frac{\sqrt{\rho^*_\text{L}} v^*_\text{L} + \sqrt{\rho^*_\text{R} } v^*_\text{R} + \left(B^*_{y\text{R}} - B^*_{y\text{L}}\right) \text{sign}(B_x)/\mu_0 }{\sqrt{\rho^*_\text{L} } + \sqrt{\rho^*_\text{R}}} \, , \\
  v^{**}_z &= \frac{\sqrt{\rho^*_\text{L} } w^*_\text{L} + \sqrt{\rho^*_\text{R} } v^*_{z\text{R}} + \left(B^*_{z\text{R} } - B^*_{z\text{L}}\right) \text{sign}(B_x)/\mu_0 }{\sqrt{\rho^*_\text{L} } + \sqrt{\rho^*_\text{R} }}, \\
   B^{**}_y &= \frac{\sqrt{\rho^*_\text{L} } B^*_{y\text{R} } + \sqrt{\rho^*_\text{R} } B^*_{y\text{L} } + \sqrt{\rho^*_\text{L} \rho^*_\text{R} \mu_0} \left(v^*_\text{R} - v^*_\text{L} \right) \text{sign}(B_x)}{\sqrt{\rho^*_\text{L} } + \sqrt{\rho^*_\text{R} }},   \\
   B^{**}_z &= \frac{\sqrt{\rho^*_\text{L} } B^*_{z\text{R}} + \sqrt{\rho^*_\text{R} } B^*_{z\text{L}} + \sqrt{\rho^*_\text{L} \rho^*_\text{R} \mu_0 } \left(v^*_{z\text{R}} - v^*_{z\text{L}}\right) \text{sign}(B_x)}{\sqrt{\rho^*_\text{L}} + \sqrt{\rho^*_\text{R}}} ,  \hspace{1.2 cm}  \\
  e^{**}_{\text{L}} &= e^{*}_{\text{L}} - \sqrt{\frac{\rho^{*}_{\text{L}}}{\mu_0}}\left(\mathbf{v}^{*}_{\text{L}} \cdot \mathbf{B}^{*}_{\text{L}} - \mathbf{v}^{**} \cdot \mathbf{B}^{**}\right) \, \text{sign}(B_x), \\
  e^{**}_{\text{R}} &= e^{*}_{\text{L}} + \sqrt{\frac{\rho^{*}_{\text{R}}}{\mu_0}}\left(\mathbf{v}^{*}_{\text{R}} \cdot \mathbf{B}^{*}_{\text{R}} - \mathbf{v}^{**} \cdot \mathbf{B}^{**}\right) \, \text{sign}(B_x) .
\end{align}
The expressions for the right state are analogous to those for the left state, with L replaced by R. The HLLD solver is capable of exactly resolving isolated discontinuities, which is the origin of the "D" in its name. This capability significantly enhances the accuracy of the solutions by capturing the full wave structure in MHD systems. However, this improvement comes with a slightly increased computational cost due to the additional calculations required for the two new intermediate states introduced in the Riemann fan. Despite this, the HLLD solver achieves an excellent balance between accuracy and efficiency, making it very useful to model for complex MHD problems.

\subsection{Constrained Transport Method}

It is important to complement these numerical methods with an additional scheme that enforces the divergence-free condition of the magnetic field. For this purpose we use the Constrained Transport Method (CT) developed by \cite{CT_evans,CT_Balsara}, which is highly compatible with the Riemann solvers described above. 
The method consists of a special discretization of the equation of Faraday. First we need to write the equation in terms of a vector ${\bf \Omega}  = {\bf v}\times {\bf B}$ like
\begin{equation}
    \frac{\partial {\bf B}}{\partial t} = \nabla \times {\bf \Omega} \, .
\end{equation}
Using this equation and finite central differences, an expression is derived for the evolution of the area-averaged components of the magnetic field on each face of the numerical cell centered at $(i,j,k)$,
\begin{align}
\frac{dB^x_{i+\frac{1}{2},j,k}}{dt} &= 
\frac{\Omega^z_{i+\frac{1}{2},j+\frac{1}{2},k} - \Omega^z_{i+\frac{1}{2},j-\frac{1}{2},k}}{\Delta y} 
- \frac{\Omega^y_{i+\frac{1}{2},j,k+\frac{1}{2}} - \Omega^y_{i+\frac{1}{2},j,k-\frac{1}{2}}}{\Delta z} \, , \\
\frac{dB^y_{i,j+\frac{1}{2},k}}{dt} &= 
\frac{\Omega^x_{i,j+\frac{1}{2},k+\frac{1}{2}} - \Omega^x_{i,j+\frac{1}{2},k-\frac{1}{2}}}{\Delta z} 
- \frac{\Omega^z_{i+\frac{1}{2},j+\frac{1}{2},k} - \Omega^z_{i-\frac{1}{2},j+\frac{1}{2},k}}{\Delta x} \, , \\
\frac{dB^z_{i,j,k+\frac{1}{2}}}{dt} &= 
\frac{\Omega^y_{i+\frac{1}{2},j,k+\frac{1}{2}} - \Omega^y_{i-\frac{1}{2},j,k+\frac{1}{2}}}{\Delta x} 
- \frac{\Omega^x_{i,j+\frac{1}{2},k+\frac{1}{2}} - \Omega^x_{i,j-\frac{1}{2},k+\frac{1}{2}}}{\Delta y} \, ,
\end{align}
where the values of $\Omega$ are calculated at the intercells as the average of the numerical fluxes, given by $F^{ij} = v^iB^j-B^iv^j$
\begin{align}
\Omega^x_{i+\frac{1}{2},j,k+\frac{1}{2}} &= 
\frac{1}{4} \left( F^{yz}_{i,j+\frac{1}{2},k} + F^{yz}_{i,j+\frac{1}{2},k+1} 
- F^{zy}_{i,j,k+\frac{1}{2}} - F^{zy}_{i+1,j,k+\frac{1}{2}} \right)  \, ,\\
\Omega^y_{i+\frac{1}{2},j,k+\frac{1}{2}} &= 
\frac{1}{4} \left( F^{zx}_{i,j,k+\frac{1}{2}} + F^{zx}_{i+1,j,k+\frac{1}{2}} 
- F^{xz}_{i+\frac{1}{2},j,k} - F^{xz}_{i+\frac{1}{2},j,k+1} \right)  \, , \\
\Omega^z_{i+\frac{1}{2},j+\frac{1}{2},k} &= 
\frac{1}{4} \left( F^{xy}_{i+\frac{1}{2},j,k} + F^{xy}_{i+\frac{1}{2},j+1,k} 
- F^{yx}_{i,j+\frac{1}{2},k} - F^{yx}_{i+1,j+\frac{1}{2},k} \right)  \, ,
\end{align}
where the fluxes $F^{ij}$ are taken from the values computed with the Riemann solver. Finally, the magnetic field in the center of the cell is obtained by the averages
\begin{align}
B^x_{i,j,k} &= \frac{1}{2} \left( B^x_{i-\frac{1}{2},j,k} + B^x_{i+\frac{1}{2},j,k} \right), \\
B^y_{i,j,k} &= \frac{1}{2} \left( B^y_{i,j-\frac{1}{2},k} + B^y_{i,j+\frac{1}{2},k} \right), \\
B^z_{i,j,k} &= \frac{1}{2} \left( B^z_{i,j,k-\frac{1}{2}} + B^z_{i,j,k+\frac{1}{2}} \right).
\end{align}
The CT method is very roboust and stable and is widely used for MHD applications including simulations of the solar atmosphere, astrophysical accretion disks, star formation, and plasma dynamics.

\end{appendix}

\end{document}